\documentclass[useAMS,usenatbib]{mn2e}
\usepackage{graphicx,color}
\usepackage{amssymb,amsmath}
\usepackage{pstricks}

\catcode`\"=\active\let"=\"
\def\pau{P. Amaro-Seoane}
\def\rainer{R. Spurzem}
\def\marc{M. Freitag}
\def\msol{{\rm M}_\odot}

\def\be{\begin{equation}}
\def\ee{\end{equation}}
\def\bc{\begin{center}}
\def\ec{\begin{center}}
\def\bfig{\begin{figure}}
\def\efig{\end{figure}}
\def\mbh{{\cal M}_{\rm bh}}
\def\mdot{\dot{\cal M}_{\rm bh}}
\def\sr{\sigma_r}
\def\s_t{\sigma_t}

\title[Accretion of stars on to a massive black hole: A realistic diffusion
model and numerical studies\\] {Accretion of stars on to a massive black hole: A
realistic diffusion model and numerical studies}

\author[\pau, \marc \, and \rainer]{\pau
  $^{1}$\thanks{E-mail: pau@ari.uni-heidelberg.de (PAS);
    freitag@ari.uni-heidelberg.de (MF);
    spurzem@ari.uni-heidelberg.de (RS)}, ~\marc$^{1}$ ~and ~\rainer$^{1}$ \\
  $^{1}$Astronomisches Rechen-Institut, M"onchhofstra{\ss}e 12-14,
  Heidelberg, D-69120, Germany\\}

\begin{document}

\date{Submitted to MNRAS on 9 January 2004}

\pagerange{\pageref{firstpage}--\pageref{lastpage}} \pubyear{2003}

\maketitle

\label{firstpage}

\begin{abstract}
  To be presented is a study of the secular evolution of a spherical
  stellar system with a central star-accreting black hole (BH) using
  the anisotropic gaseous model. This method solves numerically moment
  equations of the full Fokker-Planck equation, with Boltzmann-Vlasov
  terms on the left- and collisional terms on the right-hand sides. We
  study the growth of the central BH due to star accretion at its
  tidal radius and the feedback of this process on to the core
  collapse as well as the post-collapse evolution of the surrounding
  stellar cluster in a self-consistent manner. Diffusion in velocity
  space into the loss-cone is approximated by a simple model. The
  results show that the self-regulated growth of the BH reaches a
  certain fraction of the total mass cluster and agree with other
  methods. Our approach is much faster than competing ones (Monte
  Carlo, $N$--body) and provides detailed informations about the time
  and space dependent evolution of all relevant properties of the
  system. In this work we present the method and study simple models
  (equal stellar masses, no stellar evolution or collisions).
  Nonetheless, a generalisation to include such effects is
  conceptually simple and under way.
\end{abstract}


\section{Introduction}

The quest for the source of the luminosities of $L \approx 10^{12}
{\rm L}_{\odot }$ produced on very small scales, jets and other
properties of quasars and other types of active galactic nuclei (AGN)
led in the 60's and 70's to a thorough research that hint to the
inkling of ``super-massive central objects'' harboured at their
centres. These were suggested to be the main source of such
characteristics \citep{LyndenBell67,LR71,Hills75}. Two years later,
\citealt{LyndenBell69} showed that the release of gravitational
binding energy by stellar accretion on to a SMBH could be the primary
powerhouse of an AGN \citep{LyndenBell69}. Nowadays, this idea seems
to be ensconced in extra galactic astronomy \citep{Rees84}.

A direct consequence of the paradigm of SMBHs at the centre of ancient
galaxies to explain the energy emitted by quasars is that relic SMBHs
should inhabit at least a fraction of present-day galaxies
\citep{Rees90}. This conclusion was first made quantitative by
\citet{Soltan82} and has recently be revisited in more detail and in
the light of recent observations by \citet{YT02}.

In the last decade, observational evidences have been accumulating
that strongly suggest that massive BHs are indeed present at the
centre of most galaxies, with a significant spheroidal component.
Mostly thanks to the {\em HST}, the kinematics in present-day universe
of gas or stars has been measured in the central parts of tens of
nearby galaxies. In almost all cases \footnote{With, notably, the
  possible exception of M33 \citep{GebhardtEtAl01,MFJ01}}, proper
modelling of the measured motions requires the presence of a central
compact dark object with a mass of a few $10^6$ to $10^9\msol$
\citep[][and references
therein]{FPPMWJ01,GRH02,PinkneyEtAl03,Kormendy03}. Note, however, that
the conclusion that such an object is indeed a BH rather than a
cluster of smaller dark objects (like neutron stars, brown dwarves
etc) has only been reached for two galaxies. The first one is the
Milky Way itself at the centre of which the case for a $3$--$4\times
10^6 \msol$ MBH has been clinched, mostly through ground-based IR
observations of the fast orbital motions of a few stars
\citep{GhezEtAl03,SchoedelEtAl03}. The second case is NGC4258, which
passes a central Keplerian gaseous disc with $\rm{H_2O}$ MASER strong
sources allowing high resolution VLBI observations down to 0.16\,pc of
the centre \citep{Miyoshi95,Herrnstein99,MoranEtAl99}.

In any case, it is nowadays largely accepted that the central dark
object required to explain kinematics data in local active and
non-active galaxies is an MBH. The large number of galaxies surveyed
has allowed to study the demographics of the MBHs and, in particular,
look for correlations with properties of the host galaxy. The most
remarkable ones are the fact that the MBH has a mass which is roughly
about $0.1\%$ of the stellar mass of the spheroidal component of the
galaxy and that the mass of the BH, $\mbh$, correlates even more tightly
with the velocity of this component.  These facts certainly strike a
close link between the formation of the galaxy and the massive object
harboured at its centre. If one extends these relations to smaller
stellar systems, one could expect that globular clusters host
so-called {\em intermediate-mass black holes}, i.e. BHs whose mass is
in the range of $10^2$--$10^4 \msol$. After having been suggested in
the 70's to explain the x-ray sources observed in globulars clusters,
later discovered to be stellar-mass binaries, this possibility has
recently be revived by two lines of observations.  First IMBHs may
explain the ultra-luminous x-ray sources (ULXs) that are present in
regions of strong stellar formation in interacting galaxies and hence
suggesting a link with young ``super stellar clusters'' (SSC),
although ULXs are typically not found at the centre of luminous SSCs.
On IMBHs and their possible link to ULXs, see the review by
\cite{MC03}. Second, recent {\em HST} observations of the stellar
kinematics at the centre of M15 around the Milky Way and G1 around M31
have been interpreted as indications of the presence of an IMBH in
both clusters \citep{vdMEtAl02,GerssenEtAl02,GerssenEtAl03,GRH02}.
However, in the case of M15, the mass of the point masses required by
the observations is compatible with zero and $N$--body models have
been made of both clusters that lack a central IMBH but are compatible
with the observations 
\citep{BaumgardtEtAl03,BaumgardtEtAl03b}
. We
note that scenarios have been proposed that would quite naturally
explain the formation of an IMBH at the centre of a stellar cluster,
through run-away stellar collisions, provided that the relaxation time
is short enough and that very massive stars ($10^2\msol < M_{\star} <
10^4 \msol$) evolve into IMBH
\citep{EbisuzakiEtAl01,PZMcM02,RFG03,GFR04}.

The theoretical study of the structure and evolution of a stellar
cluster (galactic nucleus or globular cluster) harbouring a central
MBH started 30 years ago. However, due to the complex nature of the
problem which includes many physical processes and span a huge range
of time and length scales, our understanding of such systems is still
incomplete and, probably, subjected to revision. As in many fields of
astrophysics, analytical computations can only been applied to highly
idealised situations and only a very limited variety of numerical
methods have been developed so far that can tackle this problem.

In the present paper, we introduce a simulation method to follow the
joint evolution of a spherical star cluster with a central BH making
feasible anisotropy. This procedure is based on moments of the
Boltzmann equation with relaxation. The cluster is modelled like a
self-gravitating, conducting gas sphere, according to the methods
presented in \citet{LS91} and \citet{GS94}. These models improved
earlier gas models of \citet{Bettwieser83} and \citet{Heggie84}.  Much
as the structure of the numerical method is for the sake of
computational efficiency on account of physical accuracy, it allows
for all the most important physical ingredients that may carry out a
role in the evolution of a spherical cluster. These include, among
others, self-gravity, two-body relaxation, stellar evolution,
collisions, binary stars etc and, undoubtedly, the interaction with a
central BH and the role of a mass spectrum. The specific advantage of
the so-called ``gaseous model'' to other simulations methods (to be
presented in 2.2) is that the simulations are comparatively much
faster 
, since
they are grounded on numerical integration of a relatively small set
of partial differential equations with just one spatial variable, the
radius $r$. In addition, all quantities of interest are accessible as
smooth functions of $r$ and time and this allows one to investigate in
detail clean-cut aspects of the dynamics without being hindered by the
important numerical noise particle-based methods ($N$--body and Monte
Carlo) suffer from.

In this paper we concentrate on the simplest version of the gaseous
model which includes the interaction between a central BH and its host
cluster. In particular, we assume that all stars are sun-like, neglect
stellar evolution, direct collisions between stars and the role of
binary stars. Also, the only interaction between BH and the stellar
system is tidal disruptions (besides the BH's contribution to the
gravitational field), and we undertake that the BH stays fixed at the
centre of the cluster. While admittedly very simplified, we reckon this
idealised situation warrants consideration. First, it helps us to
establish that the gaseous model is also able to treat more complex
situations in phase space than self-gravitating star clusters, such as
those caused by loss-cone accretion and a central BH; second,
experience with the gaseous model and other methods showed us how
intricate the interplay between various physical effects can become
during the evolution of clusters and so we feel compelled to first
consider the simplest models to develop a robust understanding of the
mechanisms at play.

The structure of this paper is as follows: In the subsection 2.1, we
explain the physics of the problem. In 2.2 we introduce briefly the
various analytical and numerical methods used so far to investigate
spherical clusters with central BH and summarise their key results.
Later, in section 3, we present the gaseous model and, in particular,
how the interaction between the stars and the BH is included. The
interested reader may find in the Appendix A a more technical
description of the numerical method used to solve the equations of the
gaseous model. Section 4 is devoted to results obtained in a first set
of simulations. Finally, in 5, we draw conclusions about this first
use of this new code version and present what our future work with it
will likely consist of.

\section{loss-cone accretion on to massive BHs: The
diffusion model}

\subsection{Previous theoretical and numerical studies}

If we arrange numerical methods for stellar dynamics in order of
validity and both increasing the spatial resolution and decreasing the
required computational time, we can distinguish four general classes.
The most direct approach is the so-called $N$--body method
\citep{Aarseth99a,Aarseth99b,Spurzem99}. Monte Carlo codes are also
particle-based, but rely on the assumptions that the system is
spherically symmetric and in dynamical equilibrium and treat the
relaxation in the Fokker-Planck approximation (see 3.2)
(\citealt{FB04,FregeauEtAl03,FB01a,GS03,Giersz98,JoshiEtAl01}).
Ensuingly, we have the two-dimensional numerical direct solutions of
the Fokker-Planck equation
\citep{Takahashi97,Takahashi96,Takahashi95}, and the gaseous models.
The idea of this model goes back to \citet{HachisuEtAl78} and
\citet{LBE80}, who first proposed to treat the two-body relaxation as
a transport process like in a conducting plasma. They had been
developed further by \citet{Bettwieser83,BS84,Heggie84,HR89}. Their
present form, published in \citet{LS91,GS94,ST95} improves the
detailed form of the conductivities in order to yield high accuracy
(for comparison with $N$--body) and correct multi-mass models. This
point has been made already in \citet{Spurzem92}.

\citet{Peebles72,SL76} and especially \citet{FR76} and \citet{BW76} addressed
the problem of a stationary stellar density profile around a massive
star accreting BH. They found that, under certain conditions, the
density profile $\rho\propto r^{-{7/4}}$ is established in the region
where the BH's gravitational potential well dominates the self-gravity
of the stars \footnote{We must mention here the legwork done twelve
  years before this analysis by \citet{Gurevich64}, since he got an
  analogous solution for the distribution of electrons in the vicinity
  of a positively charged Coulomb centre.}.

The problem of a star cluster with a massive central star-accreting BH
has been widely coped with Fokker-Planck numerical models.  This
approach was useful in order to test the solidness of the method to
reproduce the $\rho \propto r^{-{7/4}}$ stationary density profile,
since loss-cone accretion disturbs such a density cusp \citep{OR78}.
The authors show that the stationary density profile follows from
their stellar-dynamical equation of heat transfer by scaling arguments
which are analogous to those given in \citet{SL76}.

\subsection{Loss-cone physics}

We can express the tidal radius in terms of the internal stellar 
structure re-writing equation (2) of \citet{AS01} (in pursuit AS01)

\be
r_{\rm t} \propto\Bigg( 
\frac{\mbh}{x_{\rm b} \pi \bar{\rho}} \Bigg)^{1/3},
\label{eq.r_tid_bind}
\ee

\noindent
where $\mbh$ denotes the mass of the central BH, $\bar{\rho}$ the mean
stellar internal density, $n$ is the polytropic index (stars are
supposed to be polytropes) and $x_{\rm b}$ is a parameter
proportional to the gravitational binding energy of the star that
describes effects of the internal stellar structure. We assume that a
star is disrupted by tidal forces when it crosses the tidal
radius. The free parameter $\epsilon_{\rm eff}$ (accretion efficiency)
determines the mass fraction of the gaseous debris being accreted on
to the central BH ($\epsilon_{\rm eff}=1$ corresponds to 100\%
efficiency).

There are two concurrent processes driving stars towards the tidal
radius; namely the {\em energy diffusion} and the {\em loss-cone
  accretion}. In the first case, stars on nearly circular orbits lose
energy by distant gravitational encounters with other stars and in the
process their orbits get closer and closer to the central BH.
The associated energy diffusion time-scale can be identified with the
local stellar-dynamical relaxation time generalised for anisotropy as
in \citet{Bettwieser83}: 

\be 
t_{\rm relax}= \frac {9}{16 \sqrt {\pi}}
\frac {\sr (\s_t^2/2)} {G^2 m_\star \rho_\star(r) \ln \Lambda}.  
\label{eq.t_relax}
\ee

\noindent
Here $\sr$, $\s_t$ are the radial and tangential velocity dispersions
(in case of isotropy $2\sr^2 = \s_t^2 $), $\rho_\star (r)$ is the mean
stellar mass density, $N$ the total particle number, $G$ the
gravitational constant, $m_\star$ the individual stellar mass and

\be
\ln \Lambda \equiv \ln\,(p_{\rm max}/p_0)=\ln\,(\gamma N) 
\label{eq.lambda}
\ee

\noindent
is the Coulomb logarithm. We set $\gamma=0.11$ \citep{GH94}. In this
expression $p_{\max}$ is an upper limit of $p$, the {\em impact
  parameter}; $p_0$ is the value of $p$ that corresponds to an
encounter of angle $\psi=\pi/4$, where $\psi= (\pi - \xi)/2$ if $\xi$
is defined to be the {\em deflection angle} of the encounter
\citep{Spitzer87}. In the vicinity of the BH ($r<r_{\rm h}$, see 
Sec.~\ref{subsec:units}), one
should be aware that $\Lambda \approx \mbh/m_{\star}$
\citep{BW76,LS77} but, for simplification, here we shall use
Eq.\,\ref{eq.lambda} (strictely speaking only valid at distances
$r>r_{\rm h}$) everywhere.

For a more detailed discussion of the energy
diffusion process and its description in the context of the moment
model see \citet{BS86}.

As regards the second process, the {\em loss-cone accretion}, stars
moving on radially elongated orbits are destroyed by tidal forces when
they enter the tidal radius $r_{\rm t}$. A star will belong to the
loss-cone when its {\em peribarathron} \footnote{This word fits quite
  well the idea of closest approach to this ``sinking'' hole for it
  has the meaning of no return.  The {\sl barathron}
  ($\beta\acute{\alpha} \rho\alpha\theta\rho o\nu$) was in the ancient
  Greece a cliff down to an unreacheable or unseen place where
  criminals were thrown.} (distance of closest approach to the BH, see
Fig.\ref{fig.peribarath}) is less than or equal to the tidal radius
$r_{\rm t}$, provided that its orbit is not disturbed by encounters.
Thus, the loss-cone can be defined as that part of stellar velocity
space at radius $r$, which is given by

\be 
|v_t| < v_{\rm lc}(r)=\frac {r_{\rm t}}{\sqrt {r^2-r_{\rm t}^2}} 
\cdot \sqrt {2[ \phi (r_{t}) - \phi (r)] + v_{r}(r)^2} 
\label{eq.v_lc}
\ee

\noindent 
(see AS01).
\noindent 
In the last formula $v_r$, $v_t$ are the radial and tangential
velocity of a star and

\be
\phi (r_{t}) - \phi (r)=\frac{G\mbh}{r_{t}} + \phi_{\star}(r_{t})-
\frac {G\mbh}{r}- \phi_{\star}(r)
\label{eq.diff_potentials}
\ee

\noindent
At distances $r \gg r_{\rm t}$ we can approximate this expression taking
into account that ${G\mbh}/{r_{\rm t}} \gg \phi(r)~{\rm and}~\phi(r_{\rm t})$,

\noindent
This means that

\be
v_{lc}(r) \approx \frac{r_{\rm t}}{r} \sqrt{ \frac{2G\mbh}{r_{\rm t}}}.
\ee

\noindent
On the other hand, 

\be
\sigma_{r}(r)= \sqrt{ \frac{G\mbh}{r_{\rm t}}} \cdot \bigg( 
\frac{r}{r_{\rm t}} \bigg)^{-1/2}.
 \label{eq.lc_app3}
\ee

\noindent
Thus, it is in fair approximation 

\be
v_{lc}(r) \approx \frac {r_{\rm t}}{r} \sqrt{\frac{2G\mbh}{r_{\rm t}}} 
\approx\sigma_{r}(r)\cdot\sqrt{{r_{\rm t}}/{r}}.
\label{eq.lc_app5}
\ee

\noindent
For a deeper analysis on loss-cone phenomena see AS01.

\bfig 
\resizebox{\hsize}{!}{\includegraphics[scale=0.75,bb=0 19 350 387,clip]{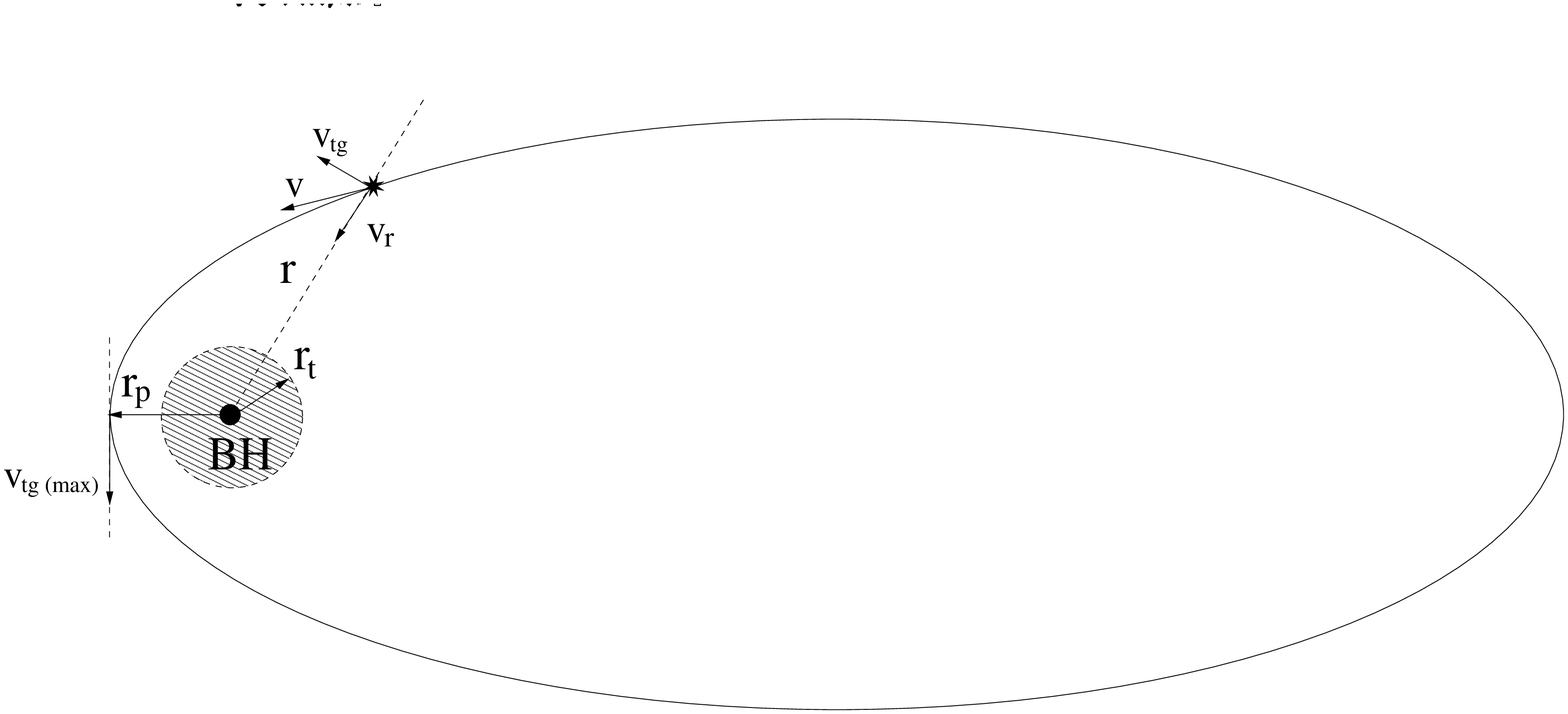}}
\caption {Definition of the {\em peribarathron} as the distance of closest 
  approximation of the star in its orbit to the BH. In this
  point the radial component of the velocity of the star cancels and
  the tangential component is maximum. In the figure ``$r_{\rm p}$'' stands
  for the peribarathron radius and ``$r_{\rm t}$'' for the tidal radius
        \label{fig.peribarath}
}   
\efig 

Similar to \citet{daCosta81}, we define two time-scales: $t_{\rm out}$
to account for the depletion of the loss-cone and $t_{\rm in}$ for its
replenishment. For a BH, $t_{\rm out}$ is equivalent to one crossing
time, since it is assumed that a star is destroyed by just one
crossing of the tidal radius. In AS01 we mentioned that in the special
case that the central object is a super-massive star, $t_{\rm
out}=n\,t_{\rm cross}$.  Here $n>1$ is the number of passages until a
star is trapped in the central object.

The loss-cone is replenished by distant gravitational encounters that
change the angular momentum vectors of the stars. To estimate $t_{\rm
  in}$, we make the assumption that random gravitational encounters
thermalise the whole velocity space at some given radius $r$ after a
time-scale of the order of $t_{\rm relax}$. As a first approximation,
the fraction $\Omega$ of the three-dimensional velocity space is
re-populated within a time-scale

\be
t_{\rm in}=\Omega \,t_{\rm relax}
\label{eq.t_in_omega}
\ee

\noindent
where 

\be
\Omega := \int_{\rm lc}f d^3v\,/\int_{\infty} f d^3v; 
\label{eq.omega}
\ee

\noindent
the subscript $\infty $ denotes an integration over the velocity space
as a whole, and the subscript $lc$ means that the integration over the
loss-cone part of the velocity space is given by Eq. (\ref{eq.v_lc}).
As a matter of fact, $\Omega$ can be envisaged as the fraction of the
surface of a velocity ellipsoid which is cut out by the loss-cone.
Close to the tidal radius $r_{\rm t}$, and for appreciable amounts of
stellar-dynamical velocity dispersion anisotropies, our method
describes the loss-cone size $\Omega $ more exactly than those given
by \citep{FR76,daCosta81}.  On the other hand, their models can be
''recovered'' in the limit of $r\gg r_{\rm t}$ and isotropy,
$2\sr^2\approx\s_t^2$ (denoted as the ``small loss-cone
approximation''), where

\be
\Omega \approx v_{\rm lc}^2/\s_t^2 \approx r_{\rm t}/r.
\label{eq.omega_app}
\ee

\noindent
This is equivalent to their definition of $\theta_{\rm lc}$ if $\Omega
= \theta_{\rm lc}^2 /4$ is adopted.
 
Where the loss-cone effects can be neglected, a
Schwarz\-schild-\-Boltz\-mann type distribution function can be
assumed,

\be
f\propto\exp\Bigl(-{(v_r-\langle v_r\rangle )^2\over
2\sr^2} -{v_t^2\over\s_t^2}\Bigr).
\label{eq.sb_DF}
\ee

\noindent
Third order moments of the velocity distribution that represent the
stellar-dynamical energy flux do not alter such a distribution
function significantly \citep{BS85}.
 
An important quantity is the {\em critical radius} $r_{\rm crit}$. Let
$\theta_{\rm D}^2$ be the average quadratic deflection angle produced
by relaxation during $t_{\rm out}$ ($=t_{\rm cross}$ here),
\be
\theta_{\rm D}^2 = \frac{t_{\rm out}}{t_{\rm relax}}.   
\ee
Then, by definition,
\be
\theta_{\rm lc}(r_{\rm crit}) = \theta_{\rm D}(r_{\rm crit}), 
\ee
which is equivalent to $t_{\rm out}(r_{\rm crit}) = 4\,t_{\rm
in}(r_{\rm crit})$.  For most clusters where such a radius can be
defined, $\theta_{\rm lc} > \theta_{\rm D}$ inside $r_{\rm crit}$
while the opposite holds outside. This means that, at large radii,
relaxation is efficient enough to make stars diffuse into and out of
loss cone orbits over a time scale $t_{\rm out}$ so that the
distribution function is not appreciably depleted in the loss
cone. Conversely, deep inside $r_{\rm crit}$, the loss cone orbits are
essentially empty and the flux of stars into this domain of phase-space
(and into the BH) can be treated as a diffusive process because the
size of one individual step of the velocity random walk process,
$\theta_{\rm D}$, is (much) smaller that the characteristic size of the
problem, $\theta_{\rm lc}$.

Note that a critical radius does not necessarily exist (see AS01). For
instance, if one assumes that gravity of the BH dominates the stellar
self-gravitation and that the density profile follows a power-law,
$\rho\propto r^{-\alpha}$, one has $\theta_{\rm lc}^2 \propto r^{-1}$,
$\theta_{\rm D}^2 \propto r^{3-\alpha}$ and a critical radius would not
exist for $x>4$.

Now we want to generalise the stationary model (see AS01), which
assumes an empty loss-cone within $r_{\rm crit}$ and a full
loss-cone elsewhere, by means of a simple ``diffusion'' model, which
is derived from the above considerations; this means that the filling
degree of the loss-cone $K$ can be continuously estimated within
its limiting values,

\be 
K \in [0,1].
\label{eq.K_limits}
\ee

\noindent
Let $f$ be the unperturbed velocity distribution (without loss-cone
accretion); if the loss-cone is empty and we neglect the angular
momentum diffusion, $f=0$ inside the loss-cone (and unchanged
elsewhere in the velocity space).  In point of fact, $f$ will have a
continuous transition from nearly unperturbed values at large angular
momenta to a partially depleted value within the loss-cone. This value
is determined by the ratio of $t_{\rm in}$ and $t_{\rm out}$.  Such a
smooth transition of the distribution function given as a function of
angular momentum $f(J)$ has been derived from self-consistent models
of angular momentum diffusion (e.g.  \citealt{CK78} or
\citealt{MS80}). We approximate $f(J)$ by a distribution function
that has a sudden jump just at the value $J_{\rm min}= m_\star v_{\rm
  lc}$ from an unperturbed value $f_0$ given by the moment equations
(assuming a Schwarzschild-Boltzmann distribution) to the constant
lowered value

\be
f=Kf_0,~{\rm with}~0\leq K \leq 1
\label{eq.f_K_f0}
\ee

\noindent
within the loss-cone (i.e. $J<J_{\rm min}$ or $\vert v_t\vert <
v_{\rm lc}$). This implies that, as means to compute the mean mass
density of loss-cone stars, we have to calculate the integral

\be
\rho_{\rm lc}=\int_{\rm lc} K \,f_0\, d^3v.
\label{eq.rho_lc}
\ee

\noindent
And then, accordant with the definition of $\Omega $,

\be 
\rho_{\rm lc, full}=\rho\,\Omega,
\label{eq.rho_omega}
\ee

\noindent
in the case that we have a full loss-cone.

\noindent
In regard to the radial and tangential stellar velocity dispersions in
the loss-cone $\sigma_{{\rm lc},\,r}$ and $\sigma_{{\rm lc},\,t}$, we can
compute them using second moments integrated over the loss-cone part
of velocity space. As for the definition of the quantities $E_r$ and
$E_t$ used in Sect. 3,
\begin{eqnarray}
\sigma_{{\rm lc},\,r}^2 &=& E_r\,\sr^2, \label{eq.E_r,t} \nonumber \\
\sigma_{{\rm lc},\,t}^2&=&E_t\,\s_t^2, 
\end{eqnarray}

\noindent
in the small loss-cone approximation we have that
$E_r\approx 1$ and $E_t \ll 1$.

The arguments about the time-scales that have led us to the derivation
of $t_{\rm in}$ and $t_{\rm out}$ guide us also to the following {\em
  diffusion equation} for the time evolution of the spatial density
$\rho_{\rm lc}=K\rho\,\Omega $ of loss-cone stars:

\be 
{d\rho_{\rm lc}\over dt}= -{\rho_{\rm lc}\,P_{\rm lc}\over t_{\rm
    out}} + \frac{\rho \,\Omega -\rho_{\rm lc}}{t_{\rm in}}.
\label{eq.diff_eq_rho_lc}
\ee

\noindent
In this equation, the second term on the right hand is the refilling term
due to relaxation.

{As we assume relaxation is due to a large number of small-angle
  deflections and can thus be seen as a diffusive process in velocity
  space,} the probability $P(\theta)$ that a star is scattered in an
angle $\theta$ in a time $t_{\rm out}$ is

\be
P({\theta})=\frac{2}{\sqrt{\pi}\,\theta_{D}}\exp\,(-{\theta^2}/
{\theta_{D}^2}),
\label{eq.P_theta}
\ee

\noindent
The distribution is normalised to one,

\be
\int^{\infty}_{0}P(\theta)\,d\theta=1
\label{eq.normP}
\ee

\noindent 
and has the property that its mean square value is
$\theta_{D}^2$. A star remains in the loss-cone during a time $t_{\rm
  out}$ if its RMS diffusion angle is smaller than the loss-cone
angle $\theta_{lc}$. The probability for this to happen is

\be
P_{lc}=\int^{\theta_{lc}}_{0}P(\theta)\,d\theta={\rm erf}(\sqrt{4\,{
t_{\rm in}}/{t_{\rm out}}}).
\label{eq.Plc}
\ee

In the case that a star is unperturbed by the rest of the stellar
system, it will sink on to the central BH in a time $t_{\rm out}$.
Actually, this is a somehow simplified description of the physical
process, for part of the loss-cone stars will be scattered out of it
before they slump. The required time for this event is $t_{\rm in}$,
since in this time-scale the angular momentum vector will change
(due to distant encounters) on an amount that is comparable with the
size of the loss-cone in the angular momentum space. For this reason
we have introduced the quantity $P_{\rm lc}$ in Eq.
(\ref{eq.diff_eq_rho_lc}).

Bluntly speaking, the {\em effective} time-scale that describes the
loss-cone depletion allowing for perturbation due to angular
momentum diffusion is

\be
t_{\rm out,\,eff} = t_{\rm out} / P_{\rm lc}.
\label{eq.t_out,eff}
\ee

\noindent
As a matter of fact, this definition ensures us that far outside of
the critical radius the loss-cone depletes in a time that grows
infinitely, as it is physically expected. In the regime where $r\ll
r_{\rm crit}$, $P_{\rm lc}$ tends asymptotically to 1 and to 0 where
$r\gg r_{\rm crit}$, passing through a transition zone at $r=r_{\rm
  crit}$.

We can consider Eq. (\ref{eq.diff_eq_rho_lc}) as an ordinary
differential equation for $K=\rho_{\rm lc}/(\rho\,\Omega)$ if we assume
that the stellar density and the loss-cone size are time-independent.
Transport phenomena can be neglected, for they are related to the
relaxation time, and $t_{\rm relax}\gg t_{\rm in},t_{\rm out}$.
Bearing this in mind we can get an analytical solution $K(t)$ for the
differential equation with the initial condition that
$K(t)|_{t_0}=K_0$,

\begin{eqnarray}
\lefteqn{K(t)=K_0\exp\Bigl(-{P_{\rm lc}\,\xi\,(t-t_0)\over t_{\rm out}}
\Bigr)+}\label{eq.K(t)} \nonumber \\
& & {} \allowbreak {t_{\rm out}\over P_{\rm lc}\,t_{\rm in}\,\xi}\cdot
\bigg( 1-\exp\Big(-{P_{\rm lc}\,\xi\,(t-t_0)\over t_{\rm
    out}}\Big)\bigg).
\end{eqnarray} 

\noindent
In the last equation we have defined $\xi$ for legibility reasons as
follows,

\be
\xi:=1+(t_{\rm out}/P_{\rm lc}\,t_{\rm in}).
\ee

\noindent
For $r=r_{\rm crit}$, with $t_{\rm in}= t_{\rm out}$, the stationary
filling degree of the loss-cone turns out to be

\be
K_{\infty}:=\lim_{t\to\infty}K(t)={1\over 2}.
\label{eq.K_infty}
\ee

\noindent
Note that \citet{MM03} recently gave a detailed summary of loss-cone
effects. They derived expressions for non-equilibrium 
configurations. They employ a rather different treatment for the
diffusion since they tackle the problem of binary BHs scattering.

\section{The gaseous model}
\subsection{Introduction}

In this section we will introduce the fundamentals of the numerical
method we use to model our system. We give a brief description of the
mathematical basis of it and the physical idea behind it. The system
is treated as a continuum, which is only adequate for a large number
of stars and in well populated regions of the phase space. We consider
here spherical symmetry and single-mass stars. We handle relaxation in
the Fokker-Planck approximation, i.e. like a diffusive process
determined by local conditions. We make also use of the hydrodynamical
approximation; that is to say, only local moments of the velocity
dispersion are considered, not the full orbital structure. In
particular, the effect of the two-body relaxation can be modelled by a
local heat flux equation with an appropriately tailored conductivity.
Neither binaries nor stellar evolution are included at the present
work. As for the hypothesis concerning the BH, see 3.4.

\subsection{The local approximation}

We treat relaxation like the addition of a big non-correlated number
of two-body encounters. Close encounters are rare and thus we admit
that each encounter produces a very small deflection angle. Thence,
relaxation can be regarded as a diffusion process \footnote{Anyhow, it
  has been argued that rare deflections with a large angle may play a
  important role in the vicinity of a BH \citep{LT80}.}.

A typical two-body encounter in a large stellar system takes place in
a volume whose linear dimensions are small compared to other typical
radii of the system (total system dimension, or scaling radii of
changes in density or velocity dispersion). Consequently, it is
assumed that an encounter only changes the velocity, not the position
of a particle.  Thenceforth, encounters do not produce any changes
${\bf\Delta x}$, so all related terms in the Fokker-Planck equation
vanish. However, the local approximation goes even further and assumes
that the entire cumulative effect of all encounters on a test particle
can approximately be calculated as if the particle were surrounded by
a very big homogeneous system with the local distribution function
(density, velocity dispersions) everywhere. We are left with a
Fokker-Planck equation containing only derivatives with respect to the
velocity variables, but still depending on the spatial coordinates (a
local Fokker-Planck equation). 

\subsection{A numerical anisotropic model}

For our description we use polar coordinates, $r$ $\theta $, $\phi$.
The vector ${\bf v} = (v_i), i=r,\theta,\phi $ denotes the velocity in
a local Cartesian coordinate system at the spatial point
$r,\theta,\phi$. For succinctness, we shall employ the notation
$u=v_r$, $v=v_\theta$, $w=v_\phi$. The distribution function $f$, is a
function of $r$, $t$, $u$, $v^2+w^2$ only due to spherical symmetry,
and is normalised according to

\be
\rho(r,t) = \int f(r,u,v^2+w^2,t) du\,dv\,dw.  
\label{eq.rho(r,t)}
\ee

\noindent
Here $\rho(r,t)$ is the mass density; if $m_{\star}$ denotes the
stellar mass, we get the particle density $n=\rho/m_{\star}$.  The
Euler-Lagrange equations of motion corresponding to the Lagrange
function

\be
{\cal L} = {1\over 2}\bigl({\dot r}^2 + r^2{\dot\theta}^2 +
         r^2 \sin^2\!\!\theta\, {\dot\phi}^2\bigr) - \Phi(r,t)
\label{eq.lagrange}
\ee

\noindent
are the following
\begin{eqnarray}
&{\dot u} =& - \frac{\partial \Phi}{\partial r} 
+ {v^2\!+\!w^2\over r} \nonumber \\
&{\dot v}  =& - {uv\over r} + {w^2\over r\tan\theta} \\
\label{eq.u_v_w_dot}
&{\dot w}  =& - {uw\over r} - {vw\over r\tan\theta} \nonumber 
\end{eqnarray}

\noindent
And so we get a complete local Fokker-Planck equation, 

\be
\frac{\partial f}{\partial t}+v_{r}\frac{\partial f}{\partial r}+\dot{v_{r}}
\frac{\partial f}{\partial v_{r}}+\dot{v_{\theta}}\frac{\partial f}{\partial 
v_{\theta}}+\dot{v_{\varphi}}\frac{\partial f}{\partial v_{\varphi}}=\Bigg
( \frac{\delta f}{\delta t} \Bigg)_{FP}
\label{eq.FP}
\ee

In our model we do not solve the equation directly; we use a so-called
{\em momenta process}. The momenta of the velocity distribution
function $f$ are defined as follows 

\be
<i,j,k>:=\int^{+\infty}_{-\infty} v^{i}_{r} v^{j}_{\theta} v^{k}_{\phi} 
\,f(r, v_{r}, v_{\theta},v_{\phi},t)\,dv_{r}dv_{\theta}dv_{\phi};
\label{eq.ijk}
\ee

\noindent
We define now the following moments of the velocity distribution function,

\begin{eqnarray}
& <0,0,0> := & \rho = \int f dudvdw \nonumber \\
& <1,0,0> := & {u} = \int uf dudvdw \nonumber \\
& <2,0,0> := & p_r + \rho {u}^2 = \int u^2 f dudvdw \nonumber \\
& <0,2,0> := & p_\theta = \int v^2 f dudvdw
\label{eq.vel_momenta} \\
& <0,0,2> := & p_\phi = \int w^2 f dudvdw \nonumber \\
& <3,0,0> := & F_r + 3{u}p_r + {u}^3 = \int u^3 f dudvdw \nonumber \\
& <1,2,0> := & F_\theta + {u}p_\theta = \int uv^2 f dudvdw \nonumber \\
& <1,0,2> := & F_\phi + {u}p_\phi = \int uw^2 f dudvdw, \nonumber
\end{eqnarray}

\noindent
where $\rho$ is the density of stars, $u$ is the bulk velocity,
$v_{r}$ and $v_{t}$ are the radial and tangential flux velocities,
$p_{r}$ and $p_{t}$ are the radial and tangential pressures, $F_{r}$
is the radial and $F_{t}$ the tangential kinetic energy flux
\citep{LS91}. Note that the definitions of $p_i$ and $F_i$ are such
that they are proportional to the random motion of the stars. Due to
spherical symmetry, we have $p_{\theta} = p_{\phi }=: p_{t}$ and
$F_{\theta} = F_{\phi} =: F_{t}/2$. By $p_{r} = \rho\sigma_{r}^2$ and
$p{_t} = \rho\sigma_{t}^2$ the random velocity dispersions are given,
which are closely related to observable properties in stellar
clusters.

\noindent
$F = (F_r + F_t)/2$ is a radial flux of random kinetic energy. In the
notion of gas dynamics it is just an energy flux. Whereas for the
$\theta-$ and $\phi-$ components in the set of Eqs.
(\ref{eq.vel_momenta}) are equal in spherical symmetry, for the $r$
and $t$- quantities this is not true. In stellar clusters the
relaxation time is larger than the dynamical time and so any possible
difference between $p_r$ and $p_t$ may {\em survive} many dynamical
times. We shall denote such differences anisotropy. Let us define the
following velocities of energy transport:

\begin{eqnarray}
&v_r & = {F_r \over 3 p_r} + u, 
\label {eq.vel_energy_trans}\\
&v_t & = {F_t \over 2 p_t} + u. \nonumber
\end{eqnarray}

\noindent
In case of {\it weak} isotropy ($p_r$=$p_t$) $2F_r$ = $3F_t$, and thus
$v_r$ = $v_t$, i.e. the (radial) transport velocities of radial and
tangential random kinetic energy are equal.

The Fokker-Planck equation (\ref{eq.FP}) is multiplicated with various
powers of the velocity components $u$, $v$, $w$. We get so up to
second order a set of moment equations: A mass equation, a continuity
equation, an Euler equation (force) and radial and tangential energy
equations. The system of equations is closed by a phenomenological
heat flux equation for the flux of radial and tangential RMS ({\em
  root mean square}) kinetic energy, both in radial direction. The
concept is physically similar to that of \citet{LBE80}. The set of
equations is

\begin{eqnarray*}
\frac{\partial{\rho}}{\partial t} + \frac{1}{r^2}\frac{\partial}
{\partial r} (r^2u\,\rho)= 0
\end{eqnarray*}

\begin{eqnarray*}
\frac{\partial u}{\partial t}+u\,\frac{\partial u}{\partial r} + 
{GM_r\over r^2} +
{1\over\rho}\frac{\partial p_r}{\partial r} + 2\,\frac{p_r - p_t}{\rho\, r} 
= 0
\end{eqnarray*}

\begin{eqnarray}
\lefteqn{
\frac{\partial{p_r}}{\partial {t}} + \frac{1}{r^2} \frac{\partial}
{\partial r} 
(r^2 u \,p_{r})+2 \,p_{r} \frac{\partial u}{\partial r} + \frac{1}{r^2} \frac{\partial}{\partial r} (r^2 F_{r}){}} \label{eq.set_of_eqs} \\
& & {} -
\frac{2F_{t}}{r} = -\frac{4}{5} 
\frac{(2p_{r}-p_{t})}
{\lambda_A t_{\rm relax}} \nonumber  
\end{eqnarray}

\begin{eqnarray*}
\lefteqn{
\frac{\partial{p_t}}{\partial {t}} + \frac{1}{r^2} \frac{\partial}
{\partial r} (r^2 u \,p_{t})+2 \,\frac{p_{t}\,u}{r}+\frac{1}{2 r^2} \frac{\partial}{\partial r}
(r^2F_{t}){}} \\
& & {} +\frac{F_{t}}{r}
= \frac{2}{5} 
\frac{(2p_{r}-p_{t})}{\lambda_A 
t_{\rm relax}}, 
\end{eqnarray*}

\noindent
where $\lambda_A$ is a numerical constant related to the time-scale of
collisional anisotropy decay. The value chosen for it has been
discussed in comparison with direct simulations performed with the
$N$--body code \citep{GS94}. The authors find that $\lambda_A=0.1$
is the physically realistic value inside the half-mass radius for all
cases of $N$, provided that close encounters and binary activity do
not carry out an important role in the system, what is, on the other
hand, inherent to systems with a big number of particles, as this is.

\noindent
With the definition of the mass $M_r$ contained in a sphere of radius
$r$

\be
\frac{\partial M_r}{\partial r} = 4 \pi r^2 \rho,
\label{eq.Mr}
\ee

\noindent
the set of Eqs. (\ref {eq.set_of_eqs}) is equivalent to gas-dynamical
equations coupled with the equation of Poisson. To close it we need an
independent relation, for moment equations of order $n$ contain
moments of order $n\,+\,1$. For this intent we use the heat conduction
closure, a phenomenological approach obtained in an analogous way to
gas dynamics. It was used for the first time by \cite{LBE80} but
restricted to isotropy. In this approximation one assumes that heat
transport is proportional to the temperature gradient,

\be
F = -\kappa \frac{\partial T}{\partial r} = -\Lambda \frac{\partial 
\sigma ^2}{\partial r}  
\label{eq.temp_grad}
\ee

\noindent
That is the reason why such models are usually also called {\em
  conducting gas sphere models}. 

It has been argued that for the classical approach
$\Lambda\propto\bar{\lambda}^2/\tau$, one has to choose the Jeans'
length $\lambda_J^2 = \sigma ^2/(4\pi G\rho)$ and the standard
Chandrasekhar local relaxation time $t_{\rm relax}\propto \sigma
^3/\rho$ \citep{LBE80}, where $\bar{\lambda}$ is the mean free path
and $\tau$ the collisional time. In this context we obtain a
conductivity $\Lambda\propto \rho/ \sigma$. We shall consider this as
a working hypothesis. For the anisotropic model we use a mean velocity
dispersion $\sigma^2 = (\sr^2 + 2\s_t^2)/3$ for the temperature
gradient and assume $v_r = v_t$ \citep{BS86}.
\noindent
Forasmuch as, the equations we need to close our model are

\begin{eqnarray}
\lefteqn{v_r - u + \frac{\lambda}{4\pi \,G\rho\,t_{\rm relax}} 
\frac{\partial \sigma^2}{\partial r} = 0} 
\label{eq.closing_eqs} \\
\lefteqn{
v_r = v_t.}  \nonumber 
\end{eqnarray}

\subsection{Inclusion of the central BH in the system}

In this subsection we discuss the way we cope with the loss-cone in
our approach. For this aim we accept the following:
  
\vskip0.2cm
\noindent
{1.}\,The system has central ($r=0$) fixed BH 

\vskip0.2cm
\noindent
{2.}\,Stars are totally destroyed when they enter $r_{\rm t}$

\vskip0.2cm
\noindent
{3.}\,Gas is completely and immediately accreted on to the BH

\vskip0.2cm

\noindent



As regards the first point, one should mention that the role of
brownian motion of the central BH can be important; as a matter of
fact, for a cluster with core radius $R_{\rm core}$, equipartition
predicts a wandering radius $R_{\rm wan}$ of order

\be
R_{\rm wan}\approx R_{\rm core}\,\sqrt{M_{\star}/\mbh},
\label{eq.wandering_radius}
\ee

\noindent
which is larger than the tidal disruption radius for BHs less massive
than $10^9 M_{\odot}$ if the core radius is 1 pc
\citep{BW76,LT80,ChatterjeeEtAl02}. The wandering of a MBH at the
centre of a cuspy cluster has been simulated by \citet{DHM03} with a
$N$-body code allowing $N=10^6$. They find that RMS velocity of the
quickly reaches equipartition with the stars but do not comment on the
wandering radius. In Appendix~\ref{apdx:wander}, we present a simple
estimate suggesting that, in a cusp $\rho \propto r^{-\alpha}$, the
wandering radius may be much reduced, $R_{\rm wan} \propto a
(m_{\star}/\mbh)^{1/(2-\alpha)}$, where a is typical length scale for
the central parts of the cluster. For $\alpha \ge 1.5$, one would then
expect $R_{\rm wan}$ to be smaller than $R_{\rm t}$ for black holes as
light as $2000\,\msol$, but this arguments neglects the flattening of
the density profile due to loss-cone accretion. Further $N$-body
simulations are clearly required to settle the question and, in
particular, if $R_{\rm wan}$ is larger than $R_{\rm t}$, to establish
the effect of the motion of the MBH on disruption rates, which can be
either increased or decreased \citep{MT99}.

We arrogate that at any sphere of radius $r$ the transport of
loss-cone stars in the time-scale $t_{\rm out,eff}$ towards the
centre happens instantaneously compared with the time step used for 
the time evolution.

Hence, the local density ``loss'' at $r$ is 

\be 
\bigg( \frac{\delta \rho}{\rm \delta t} \bigg)_{\rm lc}=
 -{\rho_{\rm lc}P_{\rm lc}\over t_{\rm out}} ;
\label{eq.drho_dt_lc}
\ee

\noindent
with $\rho_{\rm lc}=K\Omega\rho$.

\noindent
This corresponds to a local energy {\em loss} \footnote{By {\em loss}
  we mean here transport of mass and kinetic energy toward the central
  BH, for it is lost for the stellar system.} of

\begin{eqnarray}
\bigg( \frac{\delta \rho\,\sr^2}{\delta t} \bigg)_{\rm lc}&=&-\bigg( 
\frac{\delta \rho}{\delta t}\bigg)_{\rm lc} \cdot\Big( E_r\,\sr^2+u^2 \Big) 
\nonumber \\
\bigg( \frac {\delta \rho\,\s_t^2}{\delta t}\bigg)_{\rm lc}&=&-\bigg( 
\frac{\delta \rho}
{\delta t}\bigg)_{\rm lc}E_t\,\s_t^2.
\label{eq.loc_energy_loss}
\end{eqnarray}

\noindent
$E_r$ and $E_t$ are worked out integrating over the velocity
distribution part that corresponds to the loss-cone with the
approximation $u\ll\sr$ and $v_{\rm lc}\ll\s_t$,

\begin{eqnarray}
E_r&\approx& 1 \nonumber \\
E_t&\approx& v_{\rm lc}^2/\s_t^2 \ll 1
\label{eq.Er_Et}
\end{eqnarray}

\noindent
Thereupon, the mass accretion rate of the central BH can be calculated as 

\be {\dot M}=-\epsilon_{\rm eff} \int_{r_{\rm t}}^{R_{\rm tot}} \bigg(
\frac{\delta \rho}{\delta t} \bigg)_{\rm lc} 4\pi r^2 dr.
\label{eq.m_dot}
\ee

\noindent
Here $R_{\rm tot}$ stands for the total radius of the stellar system.
The accretion efficiency has been set throughout our calculations to
$\epsilon_{\rm eff}=1$; for a discussion on different $\epsilon_{\rm
  eff}$-values see \citet{MS80}.
 
The complete set of Eqs. (\ref{eq.set_of_eqs}) including the local
accretion terms of the type $(\delta/\delta t)_{\rm lc}$ for energy
diffusion and loss-cone accretion are solved implicitly. For every
time step the mass of the BH and the filling degree $K$ of the
loss-cone are brought up to the new state of the system.  The time
step is chosen in order to keep the maximum changes of the variables
below 5\% .

\noindent
For the model calculation we have utilised for the boundary conditions
that at the outer limit, $R_{\rm tot}=10^4$\,pc, we impose $u=F=0$ and
$M_{\rm r}=M_{\rm tot}$. No stellar evaporation is allowed. At the
centre, the usual boundary conditions for the gaseous model are
$u=M_{\rm r}=F=0$ but the central point $r_1=0$ is not explicitly
used when there is a BH, for obvious reasons. Instead, one imposes
that all quantities vary as power-laws, ${\rm d}\,\ln\,x/{\rm
  d}\,\ln\,r = C^{\rm st}$ inside the first non-zero radius of the
discretisation mesh, $r_2=1.7\times 10^{-6}\,{\rm pc}$ (see Appendix
A).

\subsection{Units and useful quantities}
\label{subsec:units}

The units used in the computations correspond to the so-called
$N$-body unit system, in which $G=1$, the total initial mass of the
stellar cluster is 1 and its initial total energy is $-1/2$
\citep{Henon71a,HM86}. For the simulations presented here, the 
initial cluster structure corresponds to the Plummer model whose
density profile is $\rho(r) =
\rho_0{\left(1+\left({r}/{R_{\rm Pl}}\right)^2\right)^{-5/2}}$, where
$R_{\rm Pl}$ is the Plummer scaling length.  For such a model the
$N$-body length unit is $\mathcal{U}_\mathrm{l}
={16}/(3\pi)\,R_{\rm Pl}$.

In the situations considered here, the evolution of the cluster is
driven by 2-body relaxation. Therefore, a natural time scale is the
(initial) {\em half-mass relaxation time}. We use the definition of
\citet{Spitzer87}, 
\be
\label{eq_rel_time}
T_{\rm rh}(0) = \frac{0.138 N}{\ln \Lambda}
\left(\frac{R_{1/2}^3}{G{\cal M}_{\rm cl}}\right)^{1/2}.
\ee
For a Plummer model, the half-mass radius is $R_{1/2}
=0.769\,\mathcal{U}_\mathrm{l} = 1.305\, R_{\rm Pl}$. 
${\cal M}_{\rm cl}$ is the total stellar mass.

For a cluster containing a central BH, an important quantity is the
{\em influence radius}, enclosing the central region inside of which
the gravitational influence of the BH dominates over the self-gravity
of the stellar cluster. The usual definition is $r_{\rm
h}=G\mbh/\sigma_0^2$, where $\sigma_0$ is the velocity dispersion in
the cluster at a large distance from the BH. As the latter quantity is
only well defined for a cluster with an extended core, we use here the
alternate and approximate definition $M_{\rm r}(r_{\rm h})=\mbh$,
i.e. $r_{\rm h}$ is the radius of that encloses a total stellar mass
equal to the mass of the BH.

\section{Results}
\label{sec.results}

We study the evolution of a stellar cluster with a so-called ``seed
BH'' at its centre. We consider two possible configurations for the
stellar system; one of a total mass of $M_{\rm tot}=10^5 \msol$ and
another of $10^6 \msol$. For the initial BH mass, we have chosen $\mbh
(0)=50\msol$ and $500\msol$ and we model it as a Plummer of $R_{\rm
  Pl}=1$pc. Even though it would be more realistic to set the
efficiency parameter $\epsilon_{\rm eff}=1/2$, we choose here
$\epsilon_{\rm eff}=1$ for historical reasons. Nonetheless, here we
study additionally the influence of a variation of the stellar
structure parameter $x_b$, since it influences the tidal radius and
hence the accretion rates (see Eq. \ref{eq.r_tid_bind}). For this
intent we compare case a ($x_b=1$) with another one in which we choose
the value $x_b=2$ (case b). As regards the physical meaning, the stars
of case b have twice as much internal binding energy than case a.

The cluster evolves during its pre-collapse phase up to a maximum
central density from which the energy input due to star accretion near the
tidal radius becomes sufficient in order to halt and reverse the core
collapse. Immediately afterwards, the post-collapse evolution starts.
At the beginning of the re-expansion phase, the BH significantly grows
to several $10^3$ solar masses. Thereon, a slow further expansion and
growth of the BH follow.
 
In Fig.\,\ref{fig.mbh_dmbh_1e5}, we follow the evolution of the mass
of a central BH in a globular cluster of $10^5$ stars of $1\,{\rm
  M}_\odot$. Panel (a) shows the mass of the BH
as function of time. On panel (b), we present
the accretion rate on to the BH, i.e. its growth rate. For
$\mathcal{M}_{\rm bh}(0)=50\,{\rm M}_\odot$, the early cluster's
evolution is unaffected by the presence of the BH which starts growing
suddenly at the moment of deep core collapse, around $T\simeq
14.5\,T_{\rm rh}(0)$. In Fig.\,\ref{fig.mbh_dmbh_1e6} we follow the same
evolution for the case of a stellar cluster of $10^6$ stars.

\bfig 
\resizebox{\hsize}{!}{\includegraphics[bb=33 183 280 667,clip]{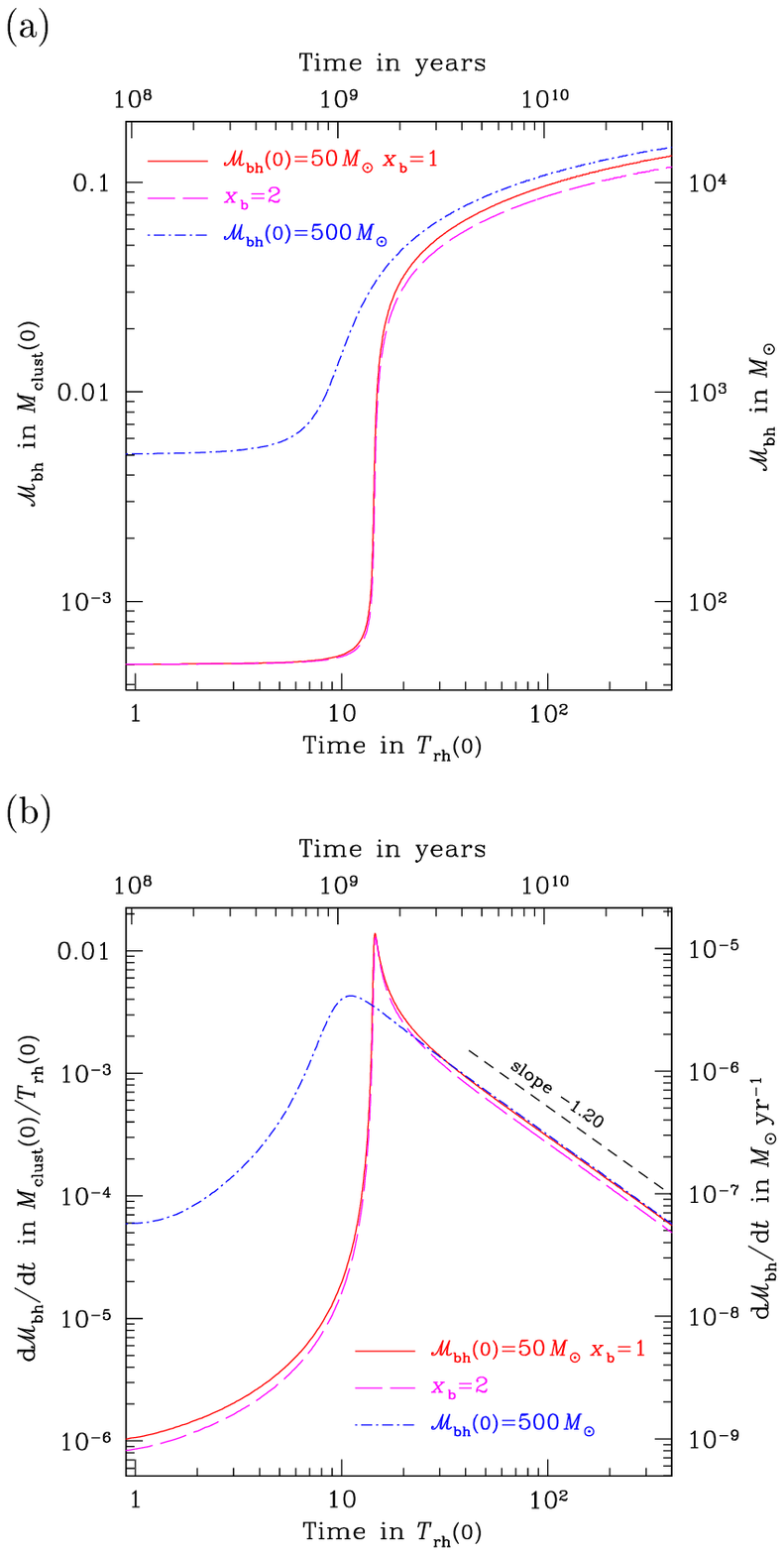}}
\caption{Evolution of the mass of a central BH in a globular cluster
  of $10^5$ stars of $1\,{\rm M}_\odot$. We considered three cases. In
  case a (solid line), the initial BH mass is $\mathcal{M}_{\rm
  bh}(0)=50\,{\rm M}_\odot$ and $x_b=1$, case b (dashes) has the same
  initial BH mass but $x_b=2$ while case c (dash-dot) corresponds to
  $\mathcal{M}_{\rm bh}(0)=500\,{\rm M}_\odot$ and $x_b=1$. An
  accretion efficiency of $\epsilon_{\rm eff}=1$ is assumed.  Panel
  (a) shows the mass of the BH as function of time and panel (b) the
  accretion rate on to the BH. At late times, the mass of the central
  BH increases like $\mdot \propto
  T^{-1.2}$ as predicted by simple scaling arguments (see text).
\label{fig.mbh_dmbh_1e5} }
\efig

From Figs.\,\ref{fig.mbh_dmbh_1e5} and \ref{fig.mbh_dmbh_1e6} we can see
that the differences between the cases a, b and c are nearly
negligible after core collapse. In general, the structure of the cluster
at late times is nearly independent of $\mbh(0)$ and $x_{\rm b}$.
From these plots we can infer that this occurs since core collapse
leads to higher densities if the initial BH mass is smaller and
thus the integrated accreted stellar mass increases.

\bfig 
\resizebox{\hsize}{!}{\includegraphics[bb=33 183 280 667,clip]{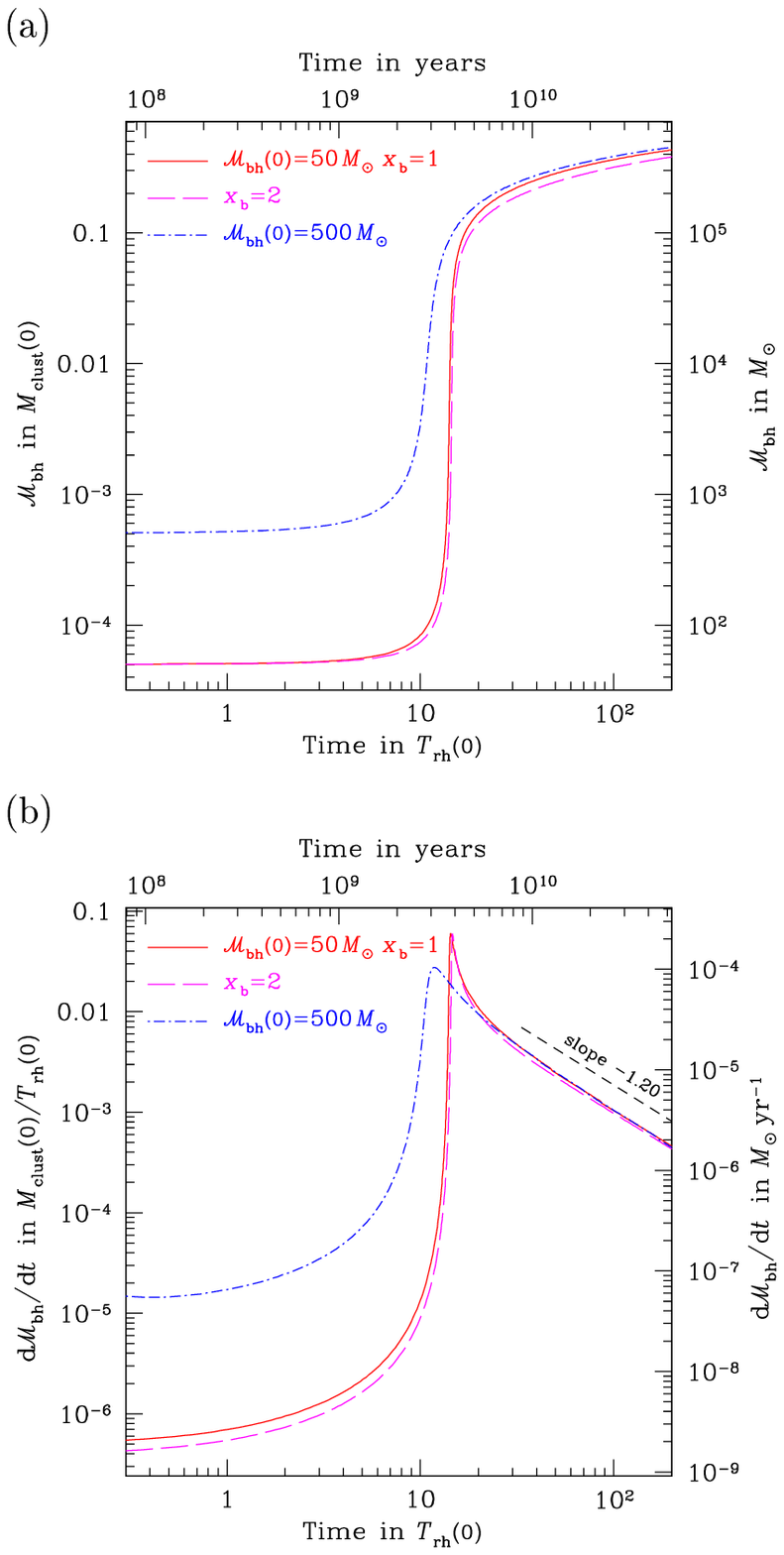}}
\caption{ Same as Fig.\,\ref{fig.mbh_dmbh_1e5} but for a cluster of $10^6$ stars 
with the same size ($R_{\rm P}=1\,{\rm pc}$).
  \label{fig.mbh_dmbh_1e6} }
\efig

We exhibit the evolution of the structure of the cluster for case a
with $10^5$ stars in Fig.\,\ref{fig.LagrRad_a1e5}. With dotted lines
we plot various Lagrangian radii, for mass fractions ranging between
$10^{-3}$\,\% (which formally corresponds to only one star) to
$90$\,\%. Only the mass still in the stellar component at a given time
is taken into account. Moreover, the evolution of the influence radius
(solid line, defined as the radius enclosing a stellar mass equal to
the BH mass) and critical radius (dashed line) are shown, so that one
can infer the percentage of the stellar mass embodied within them at a
certain moment. For late time, one obtains self-similar evolution with
size increasing like $R\propto T^{2/3}$, as expected for a system in
which the central object has a small mass and the energy production is
confined to a small central volume.
\citep{Henon65,Shapiro77,McMLC81,Goodman84}. We consider too the case
of a $10^6$ stars in Fig.\,\ref{fig.LagrRad_a1e6}, for which the
$R\propto T^{2/3}$ expansion is a poor approximation because, at late
times, the BH comprises of order 40\,\% of the system mass.

\bfig 
\resizebox{\hsize}{!}{\includegraphics[bb=34 183 572 685,clip]{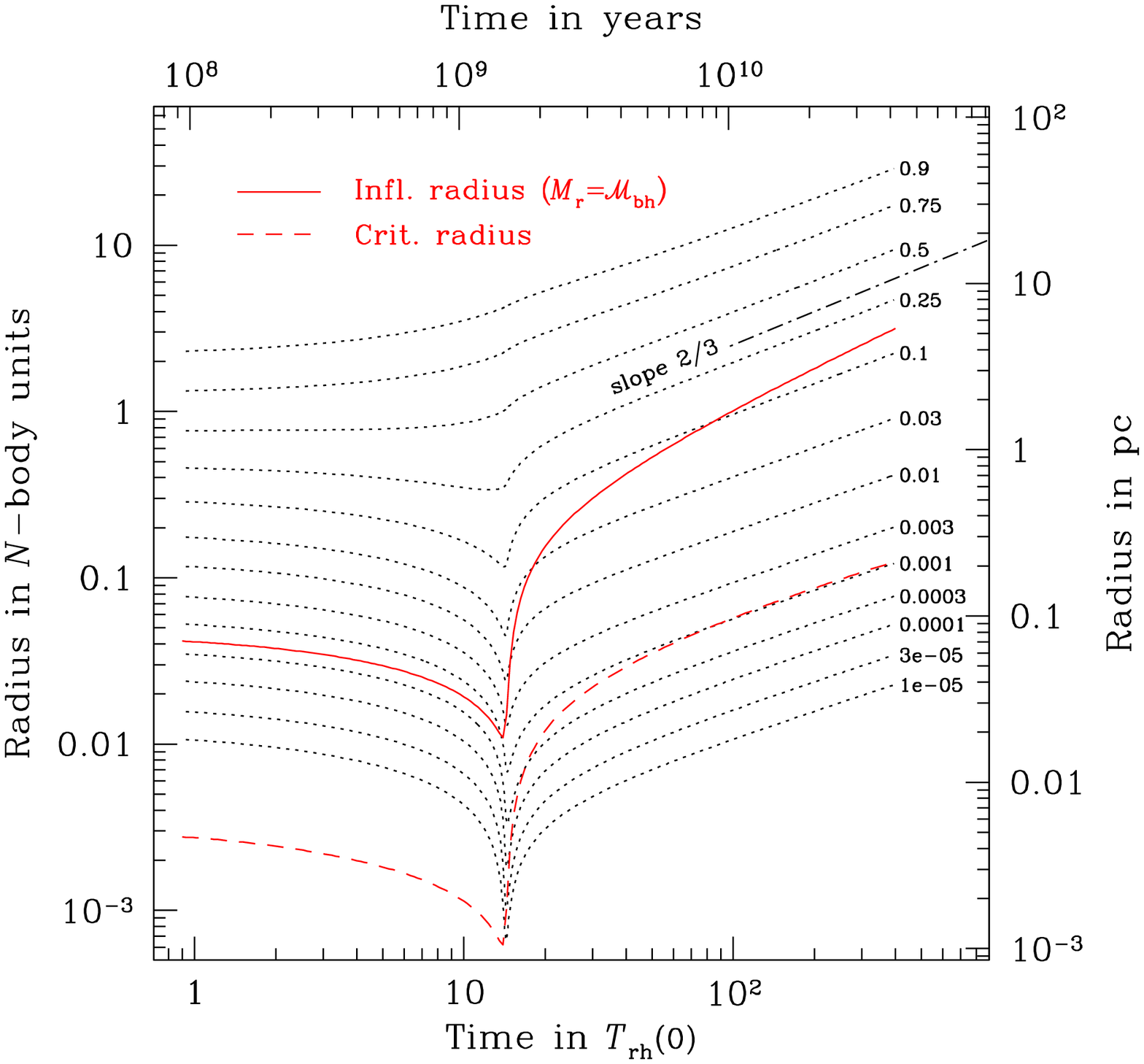}}
\caption{Evolution of the radii of spheres enclosing
  the indicated fraction of the total mass, Lagrangian radii, for case
  a. The mass fractions range from $10^{-3}$\,\% to $90$\,\%.  The
  influence and critical radius are displayed (solid and broken line).
  See text for further explanation.
\label{fig.LagrRad_a1e5} 
}
\efig

\bfig 
\resizebox{\hsize}{!}{\includegraphics[bb=34 183 572 685,clip]{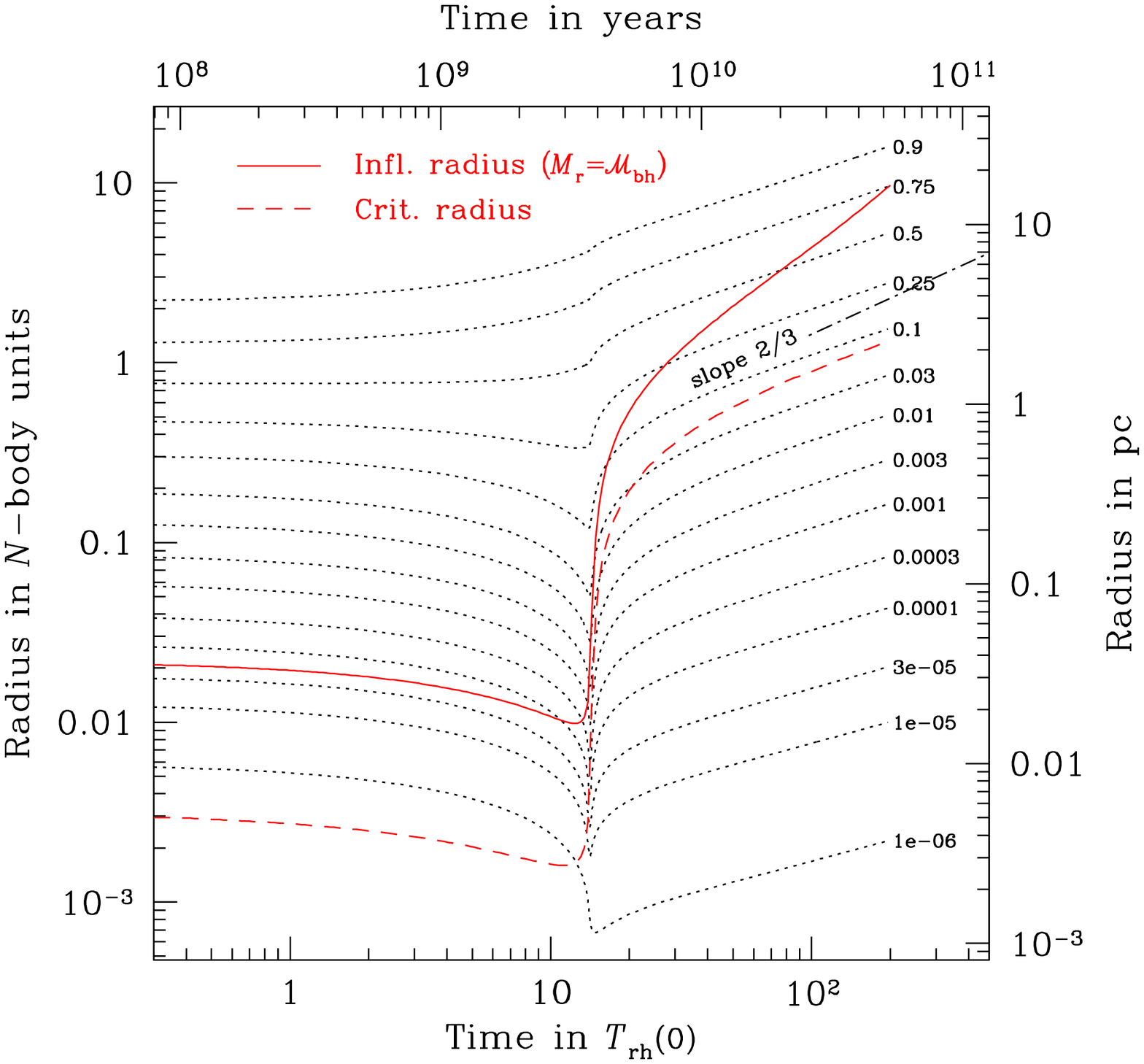}}
\caption{Same as Fig.\,\ref{fig.LagrRad_a1e5} but for a cluster of $10^6$ stars. 
 \label{fig.LagrRad_a1e6} }
\efig

We observe in Fig.\,\ref{fig.dens_a1e5_10Gyr} that for the evolved
post-collapse model the spatial profile of the stellar density has a
power law slope of $\rho\propto r^{-{7/4}}$ in the region $r_{\rm
  crit}<r<r_{\rm h}$, where $r_{\rm h}$ is the influence radius. The
density profile flattens for $r<r_{\rm crit}$ due to the effective
loss-cone accretion. For the same post-collapse moment we display the
surface density for case a in Fig.\,\ref{fig.projdens_a1e5_10Gyr}.

\noindent
\citet{LS77} proved that within $r_{\rm crit}$, according to their Eq.
(71) (where they assume $\Omega \ll 1$, small loss-cone), $\alpha$ will
continously vary from 1.75 to -$\infty$.  

\bfig 
\resizebox{\hsize}{!}{\includegraphics[bb=38 160 578 712,clip]{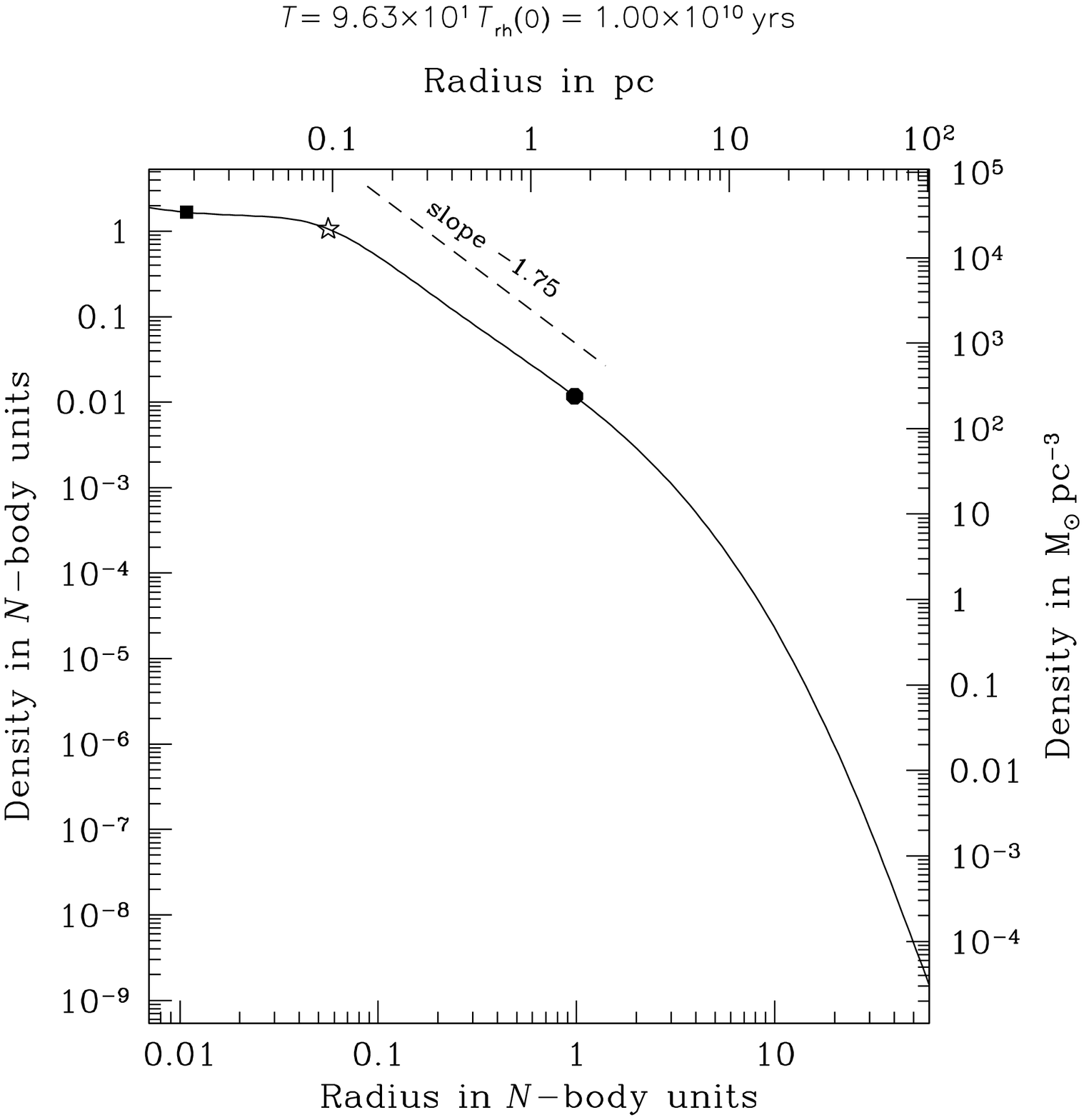}}
\caption{Density profile for a $10^5{\rm M}_\odot$ globular cluster 
  with $\mathcal{M}_{\rm bh}(0)=50\,{\rm M}_\odot$ (case a) at 10
  Gyrs. The round dot indicates the influence radius
  ($M_r=\mathcal{M}_{\rm bh}$), the star the critical radius and the
  square the radius below which the description of the cluster as a
  continuum loses significance because the enclosed mass is smaller
  than $1\,{\rm M}_\odot$. As expected, for the zone between the
  critical and influence radii, the density profile closely
  reassembles a power-law of exponent $-7/4$. At that stage, the
  structure of case c ($\mathcal{M}_{\rm bh}(0)=500\,{\rm M}_\odot$)
  is extremely similar.  
{We can see that from the ``1-star''
    radius onwards the slope of the curve shows a tendence to
    increase. This is due to the fact that the loss-cone size is
    artificially limited for stability purposes. On the other hand,
    the slope comprised between the critical radius and the 1-star
    radius is consistent with the arguments given in the work of
    \citet{LS77} (see text for further explanation).}
 \label{fig.dens_a1e5_10Gyr} 
}
\efig

\bfig 
\resizebox{\hsize}{!}{\includegraphics[bb=38 160 578 712,clip]{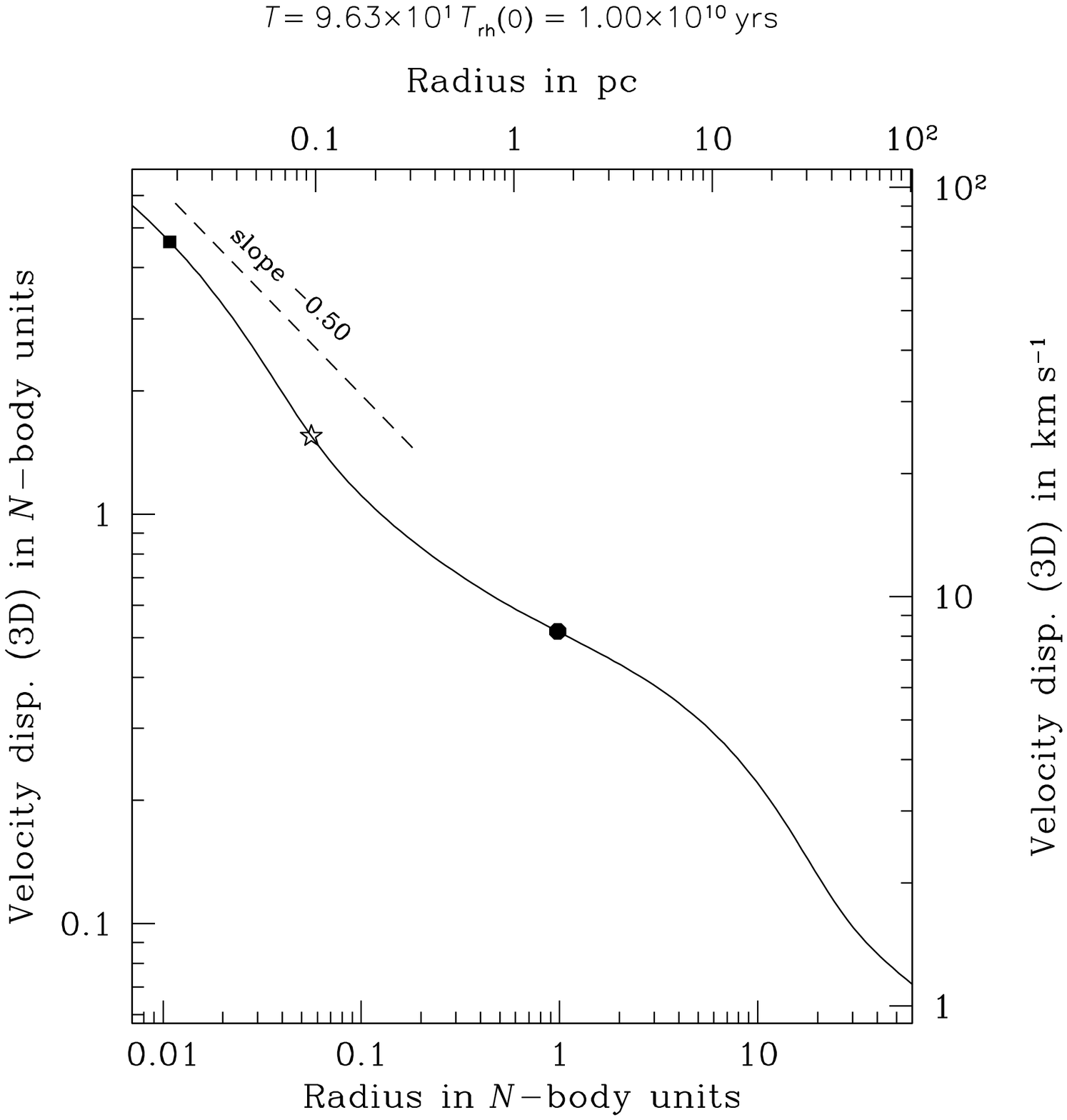}}
\caption{Profile of the three-dimensional velocity dispersion for 
the same case and time as Fig.\,\ref{fig.dens_a1e5_10Gyr}. 
{See text for comments.}   
\label{fig.disp_a1e5_10Gyr} }
\efig

\bfig 
\resizebox{\hsize}{!}{\includegraphics[bb=38 160 578 712,clip]{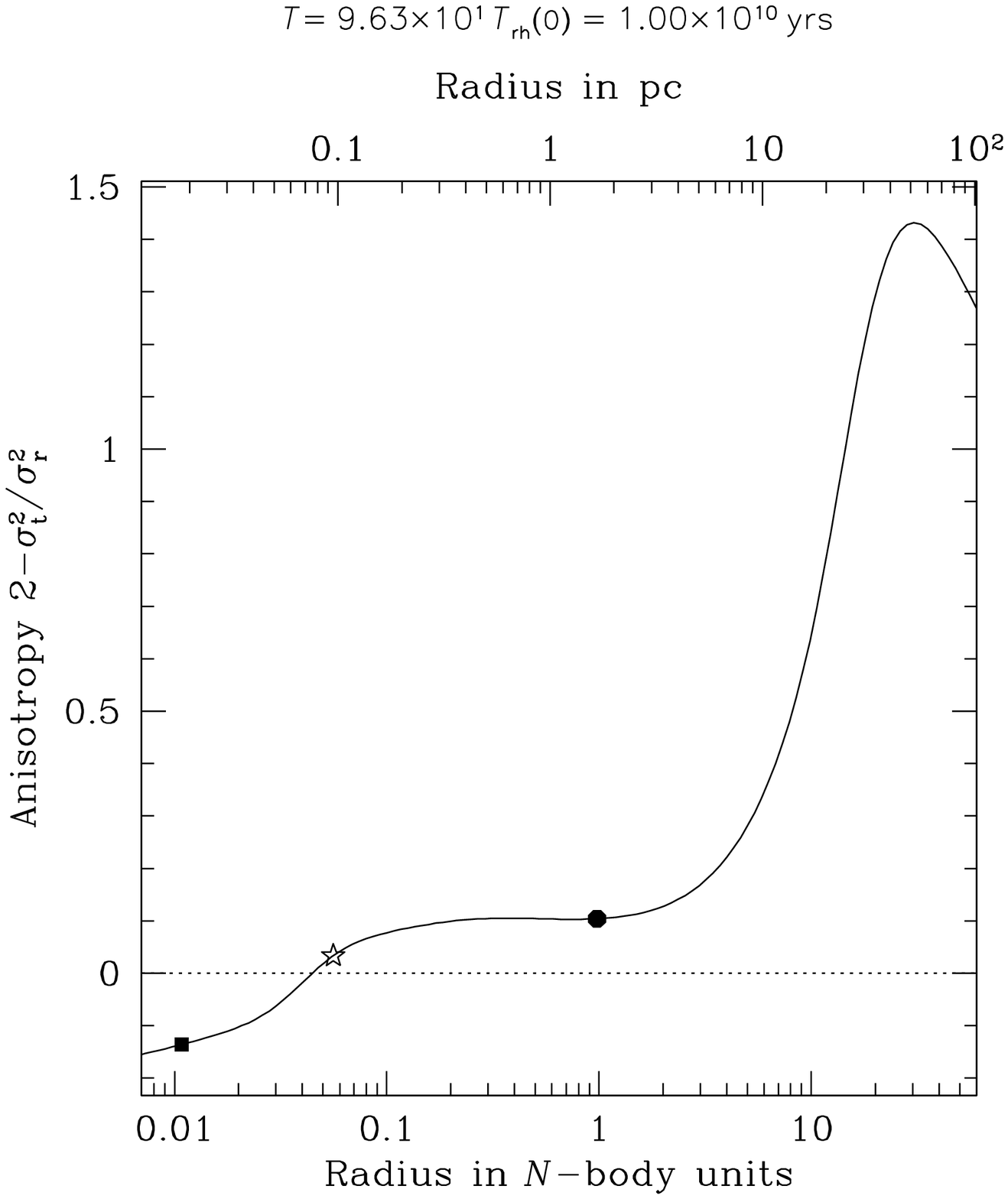}}
\caption{Profile of the anisotropy parameter for 
  the same case and time as Fig.\,\ref{fig.dens_a1e5_10Gyr}. The
  decrease of at the border is an artefact of the inappropriate
  boundary condition. An outer boundary with radial anisotropy should
  be open, but here we enforce the adiabatic wall. If one ``opens''
  the wall, but it would be at the expense of the stability of the
  program. All this does affect only a very small fraction of the
  total mass.
\label{fig.ani_a1e5_10Gyr} 
}
\efig

\bfig 
\resizebox{\hsize}{!}{\includegraphics[bb=38 160 578 712,clip]{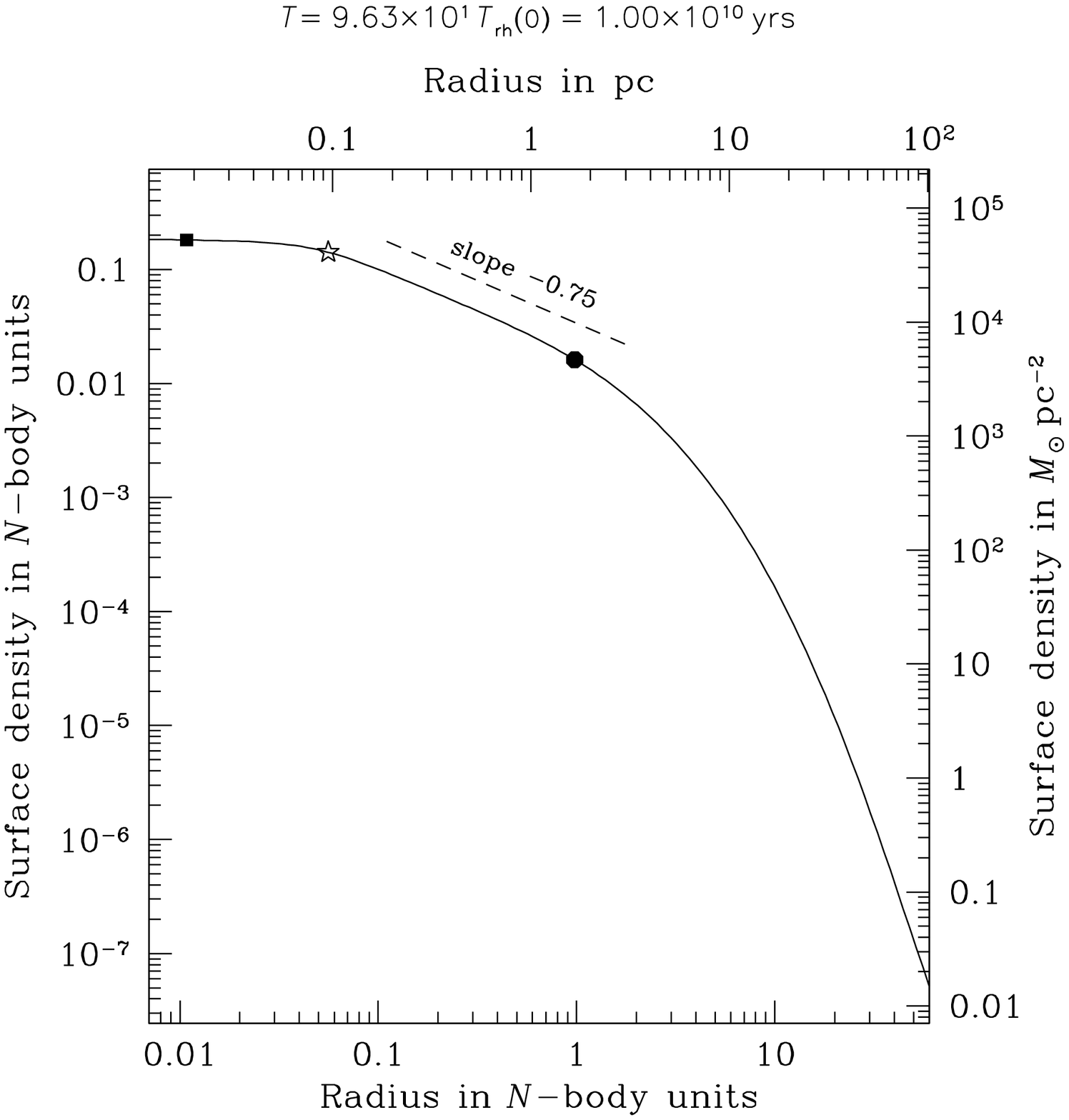}}
\caption{Projected density for the same case and time as Fig.\,\ref{fig.dens_a1e5_10Gyr}. 
  In the interval between $r_{\rm h}$ and $r_{\rm crit}$ we get a slope of
  $-3/4$, as expected.
\label{fig.projdens_a1e5_10Gyr} }
\efig

\bfig 
\resizebox{\hsize}{!}{\includegraphics[bb=38 160 578 712,clip]{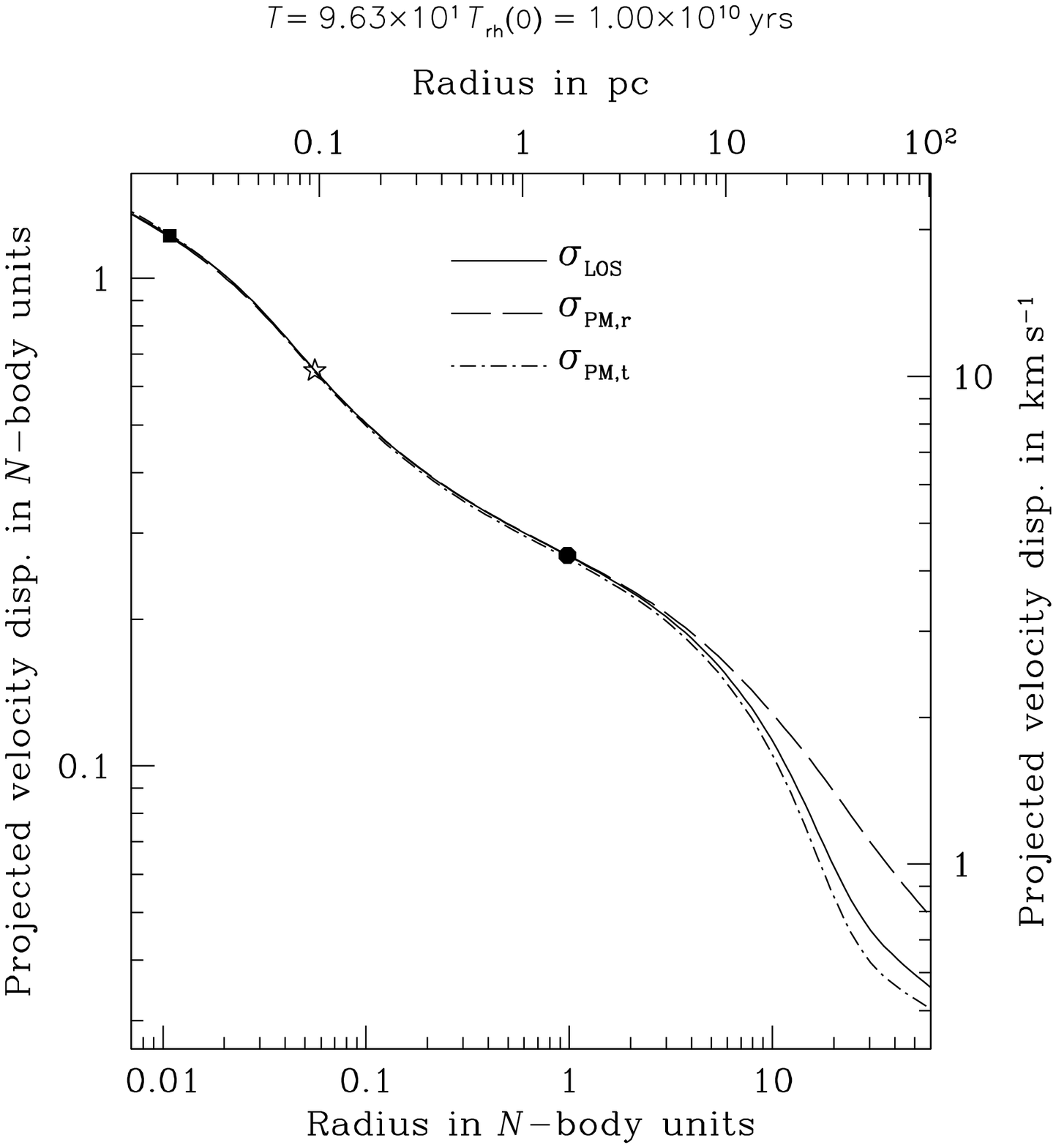}}
\caption{Projected velocity dispersions for 
the same case and time as Fig.\,\ref{fig.dens_a1e5_10Gyr}. The line of sight component is
represented with a solid line and the proper motion component with a dashed one for
the radial and dashed-dot for the tangential contribution. 
\label{fig.projdisp_a1e5_10Gyr} 
}
\efig

{ Regarding the three dimensional velocity dispersion, in
  Fig.\,\ref{fig.disp_a1e5_10Gyr} we can see coming a slope of -1/2 at
  the inner region; but extending below the one-star radius does not
  make much sense. A slope of -1/2 is what one would expect from
  Kepler's third law and a simple application of Jeans equation, with
  the assumptions that (1) dynamical equilibrium holds, (2) the
  gravity is dominated by the central BH, (3) the density follows a
  pure power-law and (4) the anisotropy $\sigma_t/\sigma_r$, is
  constant, indeed predicts $\sigma \propto r^{-1/2}$. In
  Appendix~\ref{apdx:sigma}, we show that the Jeans equation for
  stationary equilibrium actually describes the central regions of the
  cluster quite well. The reason why the velocity dispersion does not
  follow closely the ``Keplerian'' profile has to do with the fact
  that none of assumptions (2)-(4) exactly holds all the way from the
  influence radius inward.  }


Figures\,\ref{fig.projdisp_a1e5_10Gyr} and
\ref{fig.projdens_a1e5_10Gyr} give the plots of the projected density
and velocity dispersions for the late post-collapse model.

The effects of anisotropy are studied in Fig.\,\ref{fig.ani_a1e5_10Gyr},
where we can see that the external parts of the cluster are dominated
by radial orbits. Inside the critical radius (indicated by a star
symbol), one notices a slight tangential anisotropy, an effect of the
depletion of loss-cone orbits. At large radii, the velocity
distribution tends to isotropy as an effect of the outer bounding
condition imposed at $10^4$\,pc.

\noindent
The effects of anisotropy in the stellar system can also be seen in
Fig.\,\ref{fig.projdisp_a1e5_10Gyr}, where we plot the components
along the line of sight $\sigma_{\rm LOS}$ (solid line) and on the sky
(``proper motions''). The latter is decomposed into the radial
direction (i.e. towards/away from the position of the cluster's
centre) component, $\sigma_{\rm PM,r}$ (dashes) and the tangential
component, $\sigma_{\rm PM,t}$ (dash-dot). Note that the radial
anisotropy in the outskirts of the cluster reveals itself as a radial
``proper motion'' dispersion slightly larger than the other
components. For an isotropic velocity dispersion, all three components
would be equal. Despite loss-cone effects, there is no measurable
anisotropy at the centre.

The loss-cone induced anisotropy could be
detected only if one could select the stars that are known to be
spatially close to the centre (and not only in projection) as would be
feasible these stars happened to be of a particular population. An
interesting possibility that we shall soon investigate with multi-mass
models is the concentration at the centre of more massive stars,
i.e. mass segregation.

Note that some anisotropy has been detected among the stars orbiting
the central massive black hole of the Milky Way, Sgr\,A$^\ast$, at
distances closer than 1, i.e. 0.04\,pc which is well inside the
critical radius ($>1$\,pc) \citep{SchoedelEtAl03}. However, the
detected anisotropy is in the radial direction rather than
tangential. It is probably not connected to loss-cone effects but to
particular history of these seemingly very young stars \citep{GhezEtAl03}
which remains a puzzle.

\bfig 
\resizebox{\hsize}{!}{\includegraphics[bb=26 176 580 687,clip]{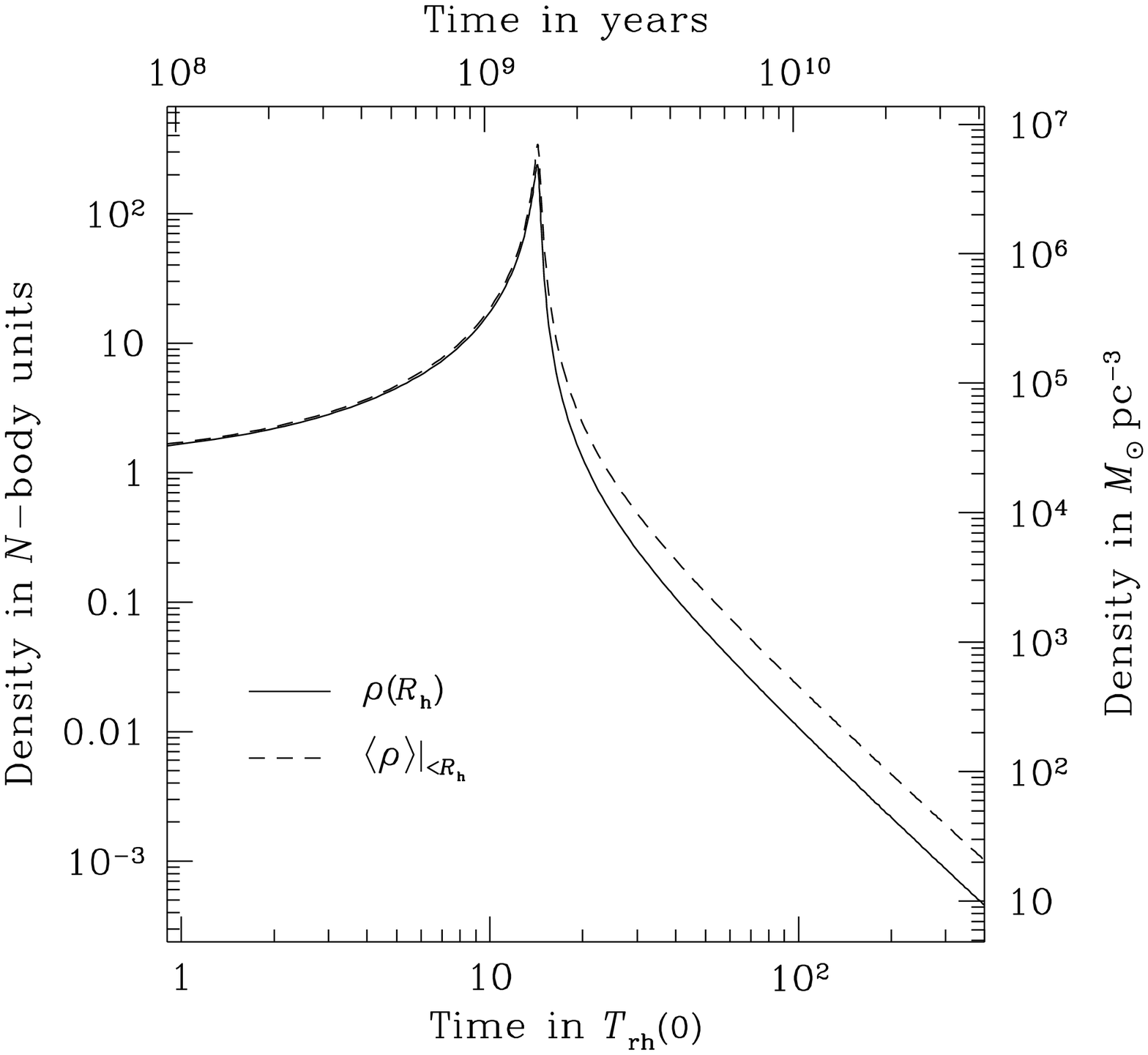}}
\caption{Evolution of the stellar density in the central region for our model with $10^5$ 
stars and $\mathcal{M}_{\rm bh}(0)=50\,{\rm M}_\odot$ (case a).  The
solid line depicts the density at the influence radius $R_{\rm
h}$. The dashed line shows the average density {\em within} $R_{\rm
h}$.
\label{fig.evol_dens_Rinfl_a1e5} }
\efig

\bfig 
\resizebox{\hsize}{!}{\includegraphics[bb=26 176 580 687,clip]{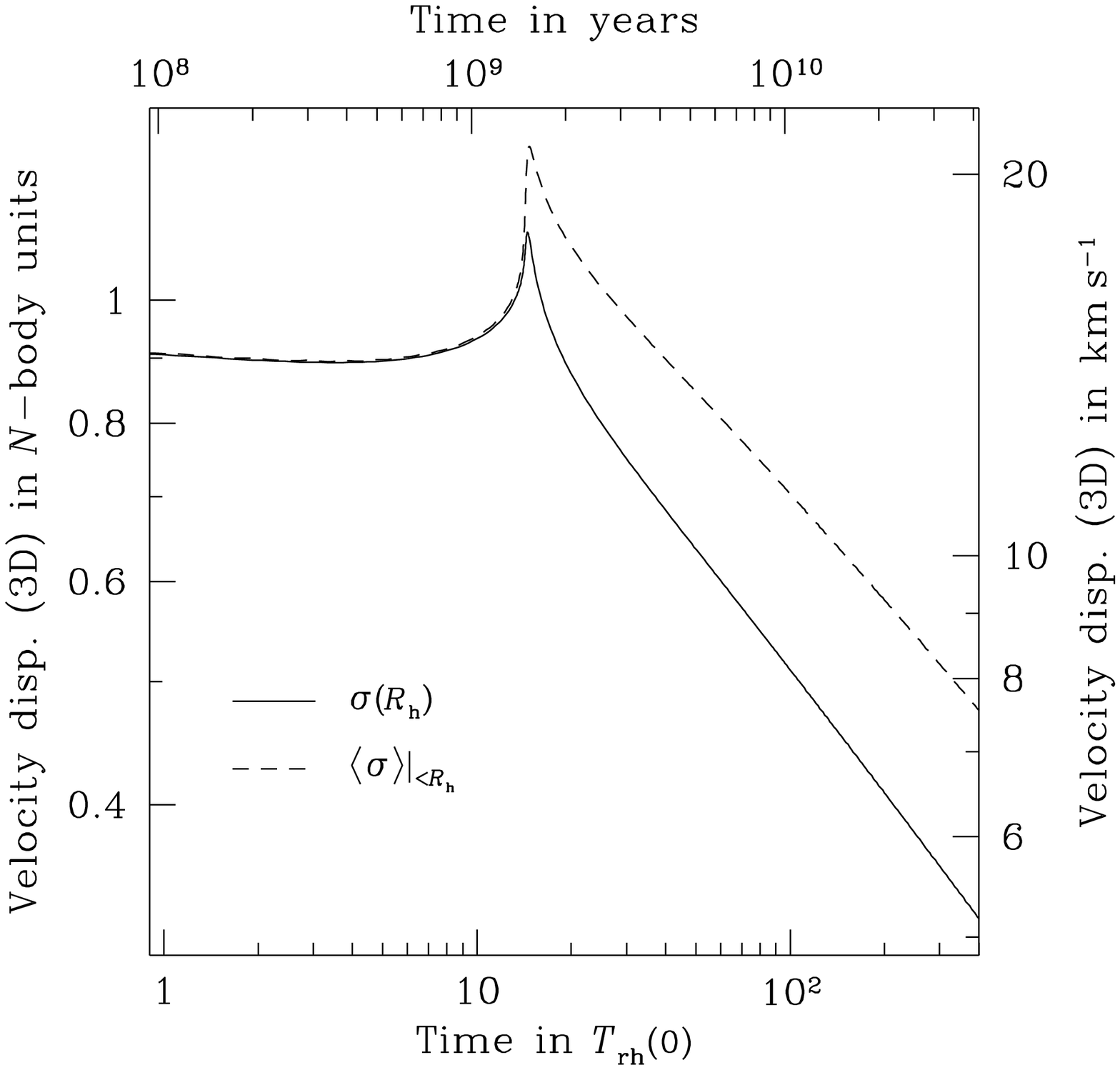}}
\caption{Same as Fig.\,\ref{fig.evol_dens_Rinfl_a1e5} but for the three dimensional velocity 
dispersion.
\label{fig.evol_disp_Rinfl_a1e5} }
\efig

In Fig.\,\ref{fig.LC_c1e5_10Gyr} the diffusion model is layed out for
the loss-cone as evaluated in previous sections. We evaluate a model
close to the post-collapse moment analysed in the other plots.  We
depict here the loss-cone filling factor $K$ (upper panel), the
loss-cone and diffusion angles $\theta_D$ and $\theta_{\rm lc}$
(middle panel) and the local contributions to the total loss-cone star
accretion rate (lower panel). Our diffusion model reproduce well the
picture of \citet{FR76}: the critical radius is defined by $\theta_D =
\theta_{\rm lc}$ and coincides with the radius where the local
contribution to the loss-cone accretion rate has its peak value.
These two angles are connected to the time-scales $t_{\rm in}$ and
$t_{\rm out}$, $\theta_D^2\propto t_{\rm out}/t_{\rm relax}$ and
$\theta_{\rm lc}^2\propto t_{\rm in}/t_{\rm relax}$. The figures show
that the maximum contribution to the mass accretion rate stems at the
radius where $\theta_D =\theta_{\rm lc}$. Consistently, \citet{FR76}
estimated the total mass accretion rate as $\mdot\propto
\rho(r_{\rm crit})\,r_{\rm crit}^3/ t_{\rm relax}(r_{\rm crit})$.

The dependence of loss-cone accretion rate on time during the late
re-expansion phase can be estimated through simple scaling laws. One
starts with the relation of \citet{FR76} mentioned above,
\be
\mdot\simeq
\frac{\rho(r_{\rm crit})\,r_{\rm crit}^3}{t_{\rm relax}(r_{\rm crit})},
\label{eq:MdotFR76}
\ee 
and the definition of the critical radius,
\be
\begin{aligned}
\theta_{\rm lc}^2(r_{\rm crit}) &= \theta_{\rm D}^2(r_{\rm crit}) \\
\Rightarrow \frac{r_{\rm t}}{r_{\rm crit}} &\simeq 
\frac{t_{\rm cross}(r_{\rm crit})}{t_{\rm relax}(r_{\rm crit})}.
\end{aligned}
\label{eq:RtoverRcrit}
\ee
One substitutes the following relations into Eq.~\ref{eq:RtoverRcrit},

\be
\begin{aligned}
r_{\rm t} &\propto \mbh^{1/3},\\
t_{\rm cross}(r) & \propto \frac{r^{3/2}}{\mbh^{1/2}}, \\
t_{\rm relax}(r) & \propto \frac{\sigma(r)^{3}}{n(r)} \propto \frac{\mbh^{3/2}}{r^{3/2}\,n(r)},
\end{aligned}
\label{eq:RelRcrit}
\ee

where we have made use of the fact that the potential is dominated by
the BH in the region of interest ($r_{\rm crit} < r_{\rm
h}$). Finally one needs the dependence of the density of stars on time
and radius, $n(r,T)$. We have seen that, to a good approximation, the
re-expansion of the cluster is homologous with Lagrange radii
expanding like $R \propto T^{2/3}$. In the region between $r_{\rm
crit}$ and $r_{\rm h}$, the density profile resembles a power-law cusp;
hence, a general self-similar evolution can be described by
\be
n(r,T) = n_0(T)\left(\frac{r}{r_0(T)}\right)^{-\alpha},
\ee
where $r_0$ is some Lagrange radius. Hence, from conservation of mass inside $r_0$,
\be
n(r,T) \propto T^{\frac{2\alpha-6}{3}}r^{-\alpha}.
\label{eq:DensOfTR}
\ee
Combining relations \ref{eq:RtoverRcrit},
\ref{eq:RelRcrit} and \ref{eq:DensOfTR}, one finds

\be
r_{\rm crit} \propto \mbh^\frac{7}{3(4-\alpha)} T^\frac{2(3-\alpha)}{3(4-\alpha)}
\ee

\noindent
and, inserting this into Eq.~\ref{eq:MdotFR76},

\be
\mdot \propto \mbh^\frac{27-19\alpha}{6(4-\alpha)} T^\frac{7(\alpha-3)}{3(4-\alpha)},
\ee

\noindent
which, by integration, yields

\begin{align}
\mbh &\propto T^\frac{2(4\alpha-9)}{13\alpha-3},\\
\mdot &\propto T^\frac{-5(\alpha-3)}{13\alpha-3}.
\end{align}

For $\alpha=7/4$, which is appropriate here (again because $r_{\rm crit} <
r_{\rm h}$), the exponent in the last relation turns out to be $-95/79
\simeq -1.20$, in remarkable agreement with figures \ref{fig.mbh_dmbh_1e5}b 
and \ref{fig.mbh_dmbh_1e6}b.

\bfig 
\resizebox{\hsize}{!}{\includegraphics[bb=104 144 529 712,clip]{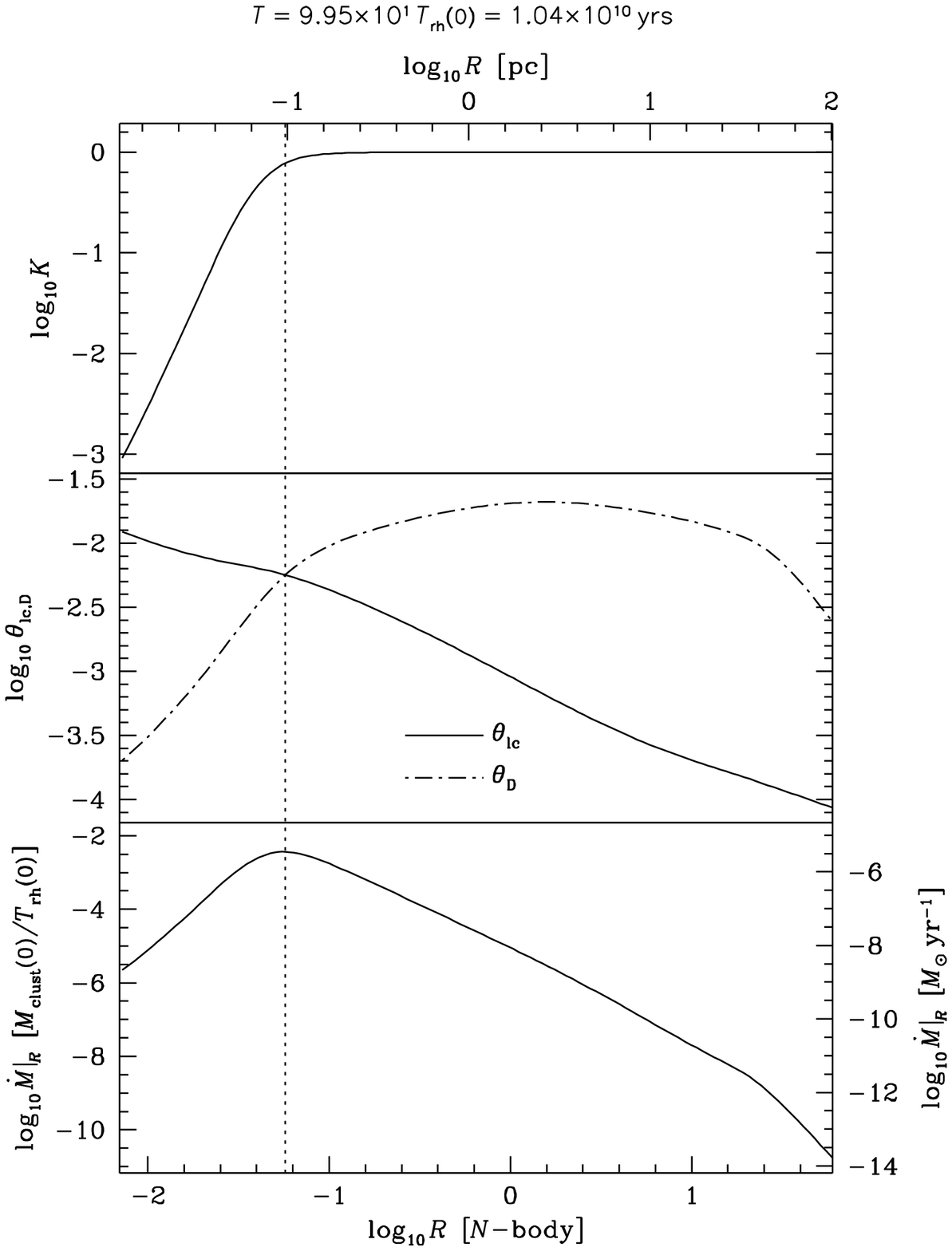}}
\caption{In this triple diagram we evince the dependence on radius of the mass 
  accretion rate, the loss-cone and diffusion angle which are related
  to the filling and depletion time-scales of the loss-cone (see text)
  and the filling factor $K$. The critical radius is defined by the
  condition $\theta_D = \theta_{\rm lc}$ (broken line).
  \label{fig.LC_c1e5_10Gyr} }
\efig

\section{Discussion}

\newcommand{\rem}[1]{} 

We have presented in this work a method to follow the evolution of a
spherical stellar cluster with a central accreting BH in a fully
self-consistent manner concerning the spatial resolution. As regards
the velocity space, we use a simplified model based on ideas of
\citet{FR76} in order to describe the behaviour of the distribution
function inside and outside the loss-cone by a simple diffusion
equation. This numerical method is an extension of the ``gaseous
model'' which has been successfully applied to a variety of aspects of
the evolution of globular clusters without central BH \citep[][amongst
others]{ST95,SA96,GS00,GS03,DS01}. With this new version, the
simulation of galactic nuclei is also feasible.

In addition to an explanation of the physical and numerical principles
underlying our approach, we have concentrated on a few simple test
computations, aimed at checking the proper behaviour of the code. We
considered a system where all stars are and remain single, have the
same mass, stellar evolution and collisions are neglected and a seed
central BH is allowed to grow by accreting stellar matter through
tidal disruptions. The present version of the code already allows for
a (discretised) stellar mass spectrum and stellar evolution and we are
in the process of including stellar collisions because they are
thought to dominate over tidal disruption in most galactic nuclei, as
far as accretion on to the BH is concerned
\citep{DDC87a,DDC87b,MCD91,FB02b}. In a subsequent paper, we shall
increase complexity and realism one step further and consider systems
with a mass spectrum. Using both this gaseous code and the Monte Carlo
algorithm \citep{FB01a,FB02b}, we will investigate the role of mass
segregation around a massive black hole (Amaro-Seoane, Freitag \&
Spurzem, in preparation), a mechanism which may have important
observational consequences as it probably affects the structure of the
central cluster of the Milky Way
\citep{Morris93,MEG00,Freitag03,Freitag03b,PL03} and impacts rates of
tidal disruptions and capture of compact stars by emission of
gravitational waves in dense galactic nuclei \citep[][and references
therein]{MT99,SU99,Sigurdsson03}.

Unfortunately, the literature has relatively little to offer to check
our models. The most robust predictions are probably the analytical
and semi-analytical analysis for the regime where the gravity of the BH
dominates and the Fokker-Planck treatment of relaxation holds
\citep{BW76,SL76,LS77,CK78}. The most important feature of these
solutions is that, provided the system is well relaxed and one stands
beyond the critical radius (inside of which loss-cone effects
complicate the picture), a cuspy density distribution is established,
$\rho\propto r^{-\alpha}$ with $\alpha=7/4$. Our code nicely agrees
with this prediction. 

Concerning the evolution of the system, we first note that, initially,
the cluster follows the usual and well understood route to
core-collapse. That the gaseous model can successfully simulate this
phase has been clearly established in previous works
\citep{GS94,SA96}. When the core has become dense enough, the BH starts 
growing quite suddenly. As it accretes stars that are deeply bound,
i.e. with very negative energies, the BH creates a outward flux of
energy and allows the cluster to re-expand. As long as the source of
energy is centrally concentrated and that the mass of the BH remains
relatively slow, one expect the re-expansion to become
self-similar, a regime during which the size of the cluster increases
like $R\propto T^{2/3}$ \citep[][among
others]{Henon65,Shapiro77,McMLC81,Goodman84}. This is again well
reproduced by the gaseous model.  Solving the Fokker-Planck equation
with a Monte Carlo method, \citet{MS80} and \citet{DS82} have realised
a series of simulations of single-mass globular clusters with a
central BH. Because their resolution was quite low and because they
used ``initial'' conditions difficult to implement (in most of their
runs the central BH is not present initially but introduced at some
instant during deep collapse), we do not attempt a quantitative
comparison with their results. An added difficulty is that we do not
include tidal truncation of the cluster. However, an important finding
of \citet{MS80} is reproduced by our computations, namely that the
initial mass of the seed black hole has little effect on the
post-collapse evolution, provided it represents only a small fraction
of cluster mass. In particular, the BH mass at late times converges to
the same value which only depends on the size and mass of the
cluster. We note that such convergence was also obtained with the
Monte Carlo algorithm and that a comparison between results obtained
with that code and an early version of the program described here was
presented in \citet{FB02b}. More comparisons between the two methods
are planned (Amaro-Seoane, Freitag \& Spurzem, in preparation).

Here one has to mention that the energy input of the BH
star accretion causes a ``temperature'' increase in the central region
which is followed by a thermal expansion. Therefore, the system is a
normal thermal system with positive specific heat in contrast to the
cores of self-gravitating systems, where the energy input due to
binary hardening causes a core expansion and decrease of temperature
\citep{BS84}. Afterwards, an inversion in the radial temperature
profile follows and the expansion is a reverse gravothermal
instability \citep{Spurzem91}. Since our system is dominated by
external gravitation in the centre we cannot expect such a
behaviour. Furthermore, as the central BH grows irreversibly, it
continues accreting stars in spite of the re-expansion and continuous
decrease of the density of stars. Hence, a second core collapse is
impossible and oscillations of the central cluster density do not
occur.

Among the aspects of our results that require further investigation,
we mention the shape of the density profile inside the critical
radius.  Although the well known $\rho \propto r^{-7/4}$ Bahcall-Wolf
solution only strictly applies for $r_{\rm crit}<r<r_{\rm h}$, this
cusp extends inward nearly down to the tidal disruption radius in the
stationary models of \cite{MS79}, which also include loss-cone
physics. This result is in disagreement with the analysis of
\citet{DO77a} (see also \citealt{OR78}) which predicted $\rho \propto
r^{-1/2}$ in this region.  As shown on Fig.~\ref{fig.dens_a1e5_10Gyr},
we obtain an even stronger flattening of the density law inside
$r_{\rm crit}$. At the present time it is unknown to us which
solution, if any, is the correct one. A possibility to be considered
is that this a consequence of the truncation of the moment equations
to the second order. In other words, in regions were the loss-cone is
significantly depleted, representing the velocity distribution by a
simple dispersion ellipsoid and using the velocity dispersion to
determine an ``effective'' loss-cone aperture (Eq.~\ref{eq.lc_app5})
is clearly quite a strong approximation. This may impact the density
distribution as the system adjusts its central structure to produce
the heating rate required by the overall expansion.

Fortunately,
numerical inaccuracies at very small radii are unlikely to affect
the overall structure and evolution of the cluster because the
loss-cone accretion physics are essentially determined by the
conditions at the critical radius, not in the immediate vicinity of
the BH.

\appendix

\section{A brief mathematical description of the code}

\newcommand{\Ngrid}{\ensuremath{N_{\rm r}}}
\newcommand{\Neq}{\ensuremath{N_{\rm eq}}}
\newcommand{\qid}[1]{\ensuremath{^{(#1)}}}
\newcommand{\rid}[1]{\ensuremath{_{#1}}}
\newcommand{\xprev}{y}
\newcommand{\myH}[1]{\tilde{#1}}
\newcommand{\GrandVect}[1]{{{\cal #1}^*}} 
\newcommand{\VecNeq}[1]{{\cal #1}}
\newcommand{\MatNeq}[1]{{\mathfrak #1}}
\newcommand{\itid}[1]{\ensuremath{_{[ #1]}}}

\newcommand{\DMatrix}{\ensuremath{{\blacksquare}}}
\newcommand{\CMatrix}{\ensuremath{{\square_{-}}}}
\newcommand{\EMatrix}{\ensuremath{{\square_{+}}}}
\newcommand{\CEMatrix}{\ensuremath{{\square_{\pm}}}}
\newcommand{\DerivMatrix}{\ensuremath{{\partial\GrandVect{F}}/{\partial \GrandVect{X}}}}
\newcommand{\DerivMatrixDisp}{\ensuremath{\frac{\partial\GrandVect{F}}{\partial \GrandVect{X}}}}

\bfig 
\resizebox{\hsize}{!}{\includegraphics[clip]{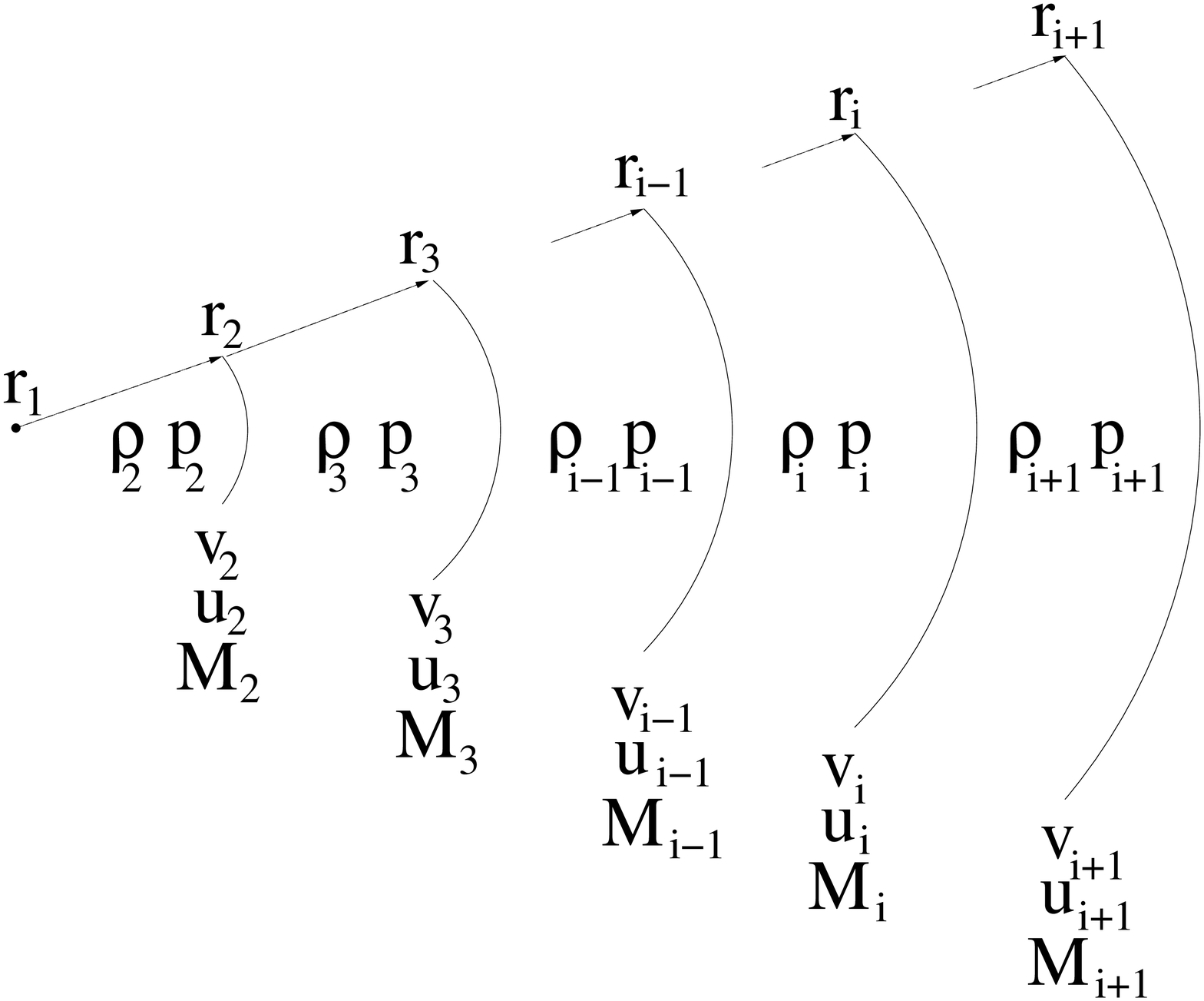}}
\caption {Representation of the logarithmic radial mesh used in the code. 
With $v$, $p$ we represent both, the radial and tangential components of the
velocity and pressure.
        \label{fig.rad_shells}
}   
\efig 

In this appendix, we explain briefly how the gaseous model is solved
numerically.  We concentrate on aspects of the method not exposed in
previous papers. This description is therefore complementary to
Sec.~2.2 of \citet{GS94}. The algorithm used is a partially implicit
Newton-Raphson-Henyey iterative scheme (\citealt{HenyeyEtAl59}, see
also \citealt{KW94}, Sec.~11.2).

Putting aside the bounding conditions, the set of equations to be
solved are Eqs.~\ref{eq.set_of_eqs} to \ref{eq.closing_eqs}. In
practice, however, the equations are rewritten using the logarithm of
all positive quantities as dependant functions. As explained in
\citet{GS94}, this greatly improves energy conservation. Formally, one
may write this system as follows
\be
\begin{split}
\frac{\partial x\qid{i}}{\partial t} + f\qid{i}\left( 
\left\{x\qid{j},\frac{\partial x\qid{j}}{\partial r}\right\}_{j=1}^{\Neq} 
\right) &=0 \mbox{\ \ for\ } i=1\dotso 4\\
f\qid{i}\left( 
\left\{x\qid{j},\frac{\partial x\qid{j}}{\partial r}\right\}_{j=1}^{\Neq} 
\right) &=0 \mbox{\ \ for\ } i=5\dotso \Neq
\end{split}
\label{eq:difeq_set}
\ee
where the $x\qid{i}$ are the local quantities defining the state of the cluster, i.e.
\be
\begin{split}
\underline{x} &\equiv \left\{x\qid{1}, x\qid{2},\dotso x\qid{\Neq}\right\} \\
& \equiv 
\left\{ \log\rho, u, \log p_r, \log p_t, \log M_r, v_r-u, v_t-u \right\},
\end{split}
\ee
with $\Neq=7$ in the present application.

To be solved numerically, this set of coupled partial differential
equations have to be discretised according to time and radius. Let us
first consider time stepping. Let $\Delta t$ be the time step. Assume
we know the solution $\underline{x}(t-\Delta t)$ at time $t-\Delta t$
and want to compute $\underline{x}(t)$. For the sake of numerical
stability, a partially implicit scheme is used. We adopt the shorthand
notations $x\qid{i} \equiv x\qid{i}(t)$ and $\xprev\qid{i} \equiv x\qid{i}(t-\Delta t)$. Time
derivation is replaced by finite differences, 
\be
\frac{\partial x\qid{i}}{\partial t} \rightarrow \Delta t^{-1}(x\qid{i}-\xprev\qid{i}).
\label{eq:time_discr}
\ee
In the terms $f\qid{i}$, we replace the $x\qid{j}$ by $\myH{x}\qid{j}$
which are values intermediate between $\xprev\qid{j}$ and $x\qid{j}$,
$\myH{x}\qid{j} =
\zeta x\qid{j} + (1-\zeta)\xprev\qid{j}$, with $\zeta= 0.55$ for
stability purpose
\citep{GS94}. 

Spatial discretisation is done by defining all quantities (at a given
time) on a radial mesh, $\{r\rid{1}, r\rid{2},\dotso r\rid{N_{\rm
r}}\}$ with $r\rid{1}=0$ and $r\rid{\Ngrid}=r_{\rm max}$. A
staggered mesh is implemented. While values of $r$, $u$, $v_t$, $v_r$
and $M_r$ are defined at the boundaries of the mesh cells, $\rho$,
$p_t$ and $p_r$ are defined at the centre of each cell, see
Fig.\,\ref{fig.rad_shells}. When the value of a ``boundary'' quantity
is needed at the centre of a cell, or vice-versa, one resort to simple
averaging, i.e. $\hat{b}\rid{k} = 0.5(b\rid{k-1}+b\rid{k})$,
$\hat{c}\rid{k} = 0.5(c\rid{k}+c\rid{k+1})$, if $b$ and $c$ are is
border- and centre-defined quantities, and $\hat{b}$, $\hat{c}$ their
centre- and border-interpolations, respectively. For all runs
presented here, $\Ngrid=300$ and the points $r\rid{2}$ to $r_{\rm
max}$ are logarithmically equidistant with $r_{\rm max}=10^4\,{\rm
pc}$ and $r\rid{2}\simeq 1.7\times 10^{-6}\,{\rm pc}$. Let us adopt the
notation $x\qid{j}\rid{k}$ for the value of $x\qid{j}$ at position $r\rid{k}$
(or $\hat{r}\rid{k}$) 
and $\Delta r\rid{k}\equiv r\rid{k}-r\rid{k-1}$. Then, radial
derivatives in the terms $f\qid{i}$ are approximated by finite differences,
\be
\frac{\partial x\qid{j}}{\partial r}
\rightarrow 
\frac{\myH{{x}}\qid{j}\rid{k}-\myH{{x}}\qid{j}\rid{k-1}}{\Delta r\rid{k}}
\label{eq:rad_discr1}
\ee
if the derivative has to be evaluated at a point where $x\rid{k}$ is
defined (centre or border of a cell), or
\be
\frac{\partial x\qid{j}}{\partial r}
\rightarrow 
\frac{\hat{\myH{{x}}}\qid{j}\rid{k}-\hat{\myH{{x}}}\qid{j}\rid{k-1}}{\Delta r\rid{k}} =
\frac{\myH{x}\qid{j}\rid{k+1}-\myH{x}\qid{j}\rid{k-1}}{2\Delta r\rid{k}}
\label{eq:rad_discr2}
\ee
otherwise. As an exception we use upstream differencing in
$\partial{u}/\partial{r}$ for the second equation in set
\ref{eq.set_of_eqs}, i.e. the difference quotient is displaced by half
a mesh point upstream to improve stability.

By making the substitutions for ${\partial x\qid{j}}/{\partial t}$ and
${\partial x\qid{j}}/{\partial r}$ in the set of differential
equations
\ref{eq:difeq_set}, one obtains, at each mesh point $r\rid{k}$, a set
of $\Neq$ non-linear algebraic equations linking the new values to be
determined, $\underline{x}\rid{k-1}$ and $\underline{x}\rid{k}$, to
the ``old'' ones, $\underline{\xprev}\rid{k-1}$ and
$\underline{\xprev}\rid{k}$, which are known,
\be
\begin{split}
 {\cal F}\qid{i}\rid{k}\left(\underline{x}\rid{k-1},\underline{x}\rid{k} |
\underline{\xprev}\rid{k-1},\underline{\xprev}\rid{k}\right)=0\\
 \ i=1\dotso \Neq, \ k=1\dotso \Ngrid.
\end{split}
\ee

Note that the structure of the equations is the same at all mesh
points, except $k=1$ and $k=\Ngrid$. In particular, terms $k-1$ do
not appear in ${\cal F}\qid{i}\rid{1}$. Also, one has to keep in mind that
only the $\underline{x}\rid{k-1}$ and $\underline{x}\rid{k}$ are
unknown; the $\underline{\xprev}\rid{k-1}$ and
$\underline{\xprev}\rid{k}$ play the role of fixed parameters in these
equation (as do the $\Delta r\rid{k}$). If one defines a $(\Neq\times
\Ngrid)$-dimension vector $\GrandVect{X}$ whose component
$\Neq(k-1)+i$ is $x\qid{i}\rid{k}$, one can write the system of $\Neq\times
\Ngrid$ equations as $\GrandVect{F}(\GrandVect{X})=0$, i.e.
\be
\GrandVect{F}(\GrandVect{X}) \equiv \left( \begin{array}{c}
{\cal F}\qid{1}\rid{1} \\
{\cal F}\qid{2}\rid{1} \\
\vdots \\
{\cal F}\qid{\Neq}\rid{1} \\
{\cal F}\qid{1}\rid{2} \\
\vdots \\
{\cal F}\qid{\Neq}\rid{2} \\
\vdots \\
{\cal F}\qid{1}\rid{\Ngrid} \\
\vdots \\
{\cal F}\qid{\Neq}\rid{\Ngrid} \\
\end{array}\right) =  \left( \begin{array}{c}
0 \\
\vdots \\
0 \\
\end{array}\right)
\mbox{\ with\ }\GrandVect{X} \equiv \left( \begin{array}{c}
x\qid{1}\rid{1} \\
x\qid{2}\rid{1} \\
\vdots \\
x\qid{\Neq}\rid{1} \\
x\qid{1}\rid{2} \\
\vdots \\
x\qid{\Neq}\rid{2} \\
\vdots \\
x\qid{1}\rid{\Ngrid} \\
\vdots \\
x\qid{\Neq}\rid{\Ngrid} \\
\end{array}\right).
\label{eq:discreq_set}
\ee

The system is solved iteratively using Newton-Raphson scheme. If
$\GrandVect{X}\itid{m}$ is the approximation to the solution of
Eq.~\ref{eq:discreq_set} after iteration $m$, with
$\GrandVect{F}\itid{m} \equiv \GrandVect{F}(\GrandVect{X}\itid{m})\ne 0$, the solution is refined
through the relation
\be
\GrandVect{X}\itid{m+1} = \GrandVect{X}\itid{m} - 
\left(\DerivMatrixDisp\right)^{-1} 
\GrandVect{F}\itid{m}
\label{eq:NewtRaphsonStep}
\ee
where $(\DerivMatrix)^{-1}$ is
the inverse of the matrix of derivatives. The latter, of dimension
$(\Neq\, \Ngrid)\times(\Neq\, \Ngrid)$, has the following structure

\be
\setcounter{MaxMatrixCols}{20}
\frac{\partial \GrandVect{F}}{\partial \GrandVect{X}} =
\begin{pmatrix}
\DMatrix & \EMatrix \\
\CMatrix & \DMatrix & \EMatrix \\
 & \CMatrix & \DMatrix & \EMatrix \\
 & & \ddots & \ddots\\
 & & \CMatrix\rid{k} & \DMatrix\rid{k} & \EMatrix\rid{k} \\
 & & & \ddots & \ddots\\
 & & & \CMatrix & \DMatrix & \EMatrix \\
 & & & & \CMatrix & \DMatrix \\
\end{pmatrix}.
\ee
\setcounter{MaxMatrixCols}{10}
In this diagram, each square is a $\Neq\times \Neq$ sub-matrix. For $2\le k
\le \Ngrid-1$, lines ${\Neq}k-6$ to ${\Neq}k$ of $\DerivMatrix$ are composed
of a group of 3 such ${\Neq}\times {\Neq}$ matrices, $\CMatrix\rid{k},
\DMatrix\rid{k}, \EMatrix\rid{k}$ that span columns ${\Neq}k-13$ to ${\Neq}k+{\Neq}$, while the rest is composed of zeros,

\be
\begin{aligned}
\DMatrix\rid{k} &= \begin{pmatrix}
\frac{\partial {\cal F}\qid{1}\rid{k}}{\partial x\qid{1}\rid{k}} & \frac{\partial {\cal F}\qid{1}\rid{k}}{\partial x\qid{2}\rid{k}} & \cdots & \frac{\partial {\cal F}\qid{1}\rid{k}}{\partial x\qid{{\Neq}}\rid{k}} \\
\vdots & & & \vdots \\
\frac{\partial {\cal F}\qid{{\Neq}}\rid{k}}{\partial x\qid{1}\rid{k}} & \frac{\partial {\cal F}\qid{{\Neq}}\rid{k}}{\partial x\qid{2}\rid{k}} & \cdots & \frac{\partial {\cal F}\qid{{\Neq}}\rid{k}}{\partial x\qid{{\Neq}}\rid{k}} \\
\end{pmatrix}\\
\CEMatrix\rid{k} &= \begin{pmatrix}
\frac{\partial {\cal F}\qid{1}\rid{k}}{\partial x\qid{1}\rid{k\pm 1}} & \frac{\partial {\cal F}\qid{1}\rid{k}}{\partial x\qid{2}\rid{k\pm 1}} & \cdots & \frac{\partial {\cal F}\qid{1}\rid{k}}{\partial x\qid{{\Neq}}\rid{k\pm 1}} \\
\vdots & & & \vdots \\
\frac{\partial {\cal F}\qid{{\Neq}}\rid{k}}{\partial x\qid{1}\rid{k\pm 1}} & \frac{\partial {\cal F}\qid{{\Neq}}\rid{k}}{\partial x\qid{2}\rid{k\pm 1}} & \cdots & \frac{\partial {\cal F}\qid{{\Neq}}\rid{k}}{\partial x\qid{{\Neq}}\rid{k\pm 1}} \\
\end{pmatrix}.\\
\end{aligned}
\ee
%

The Heyney method is a way to take advantage of the special structure
of matrix $\DerivMatrix$ to solve system \ref{eq:NewtRaphsonStep}
efficiently, with a number of operation scaling like ${\cal O}(N_{\rm
r})$ rather than ${\cal O}(\Ngrid^3)$ as would be the case if one
uses a general-purpose matrix inversion scheme\footnote{Memory usage
is also reduced, scaling like ${\cal O}(\Ngrid)$ rather than ${\cal
O}(\Ngrid^2)$.}. Setting
$\GrandVect{B}\equiv-\GrandVect{F}\itid{m}$ and $\GrandVect{W}\equiv
\GrandVect{X}\itid{m+1}-\GrandVect{X}\itid{m}$,
Eq.~\ref{eq:NewtRaphsonStep} is equivalent to
\be
  \left(\DerivMatrixDisp\right) \GrandVect{W} = \GrandVect{B}
\ee
with $\GrandVect{W}$ the unknown vector. We further decompose vectors
$\GrandVect{W}$ and $\GrandVect{B}$ into {\Neq}--dimensional sub-vectors,
each one representing the values at a given mesh points,
\be
\GrandVect{W} = \begin{pmatrix}
\VecNeq{W}\rid{1} \\
\VecNeq{W}\rid{2} \\
\vdots \\
\VecNeq{W}\rid{k} \\
\vdots \\
\VecNeq{W}\rid{\Ngrid} \\
\end{pmatrix}.
\label{eq:MtimesWequalB}
\ee
Then, the system \ref{eq:MtimesWequalB} can be written as a set of
coupled {\Neq}--dimensional vector equations,
\be
\addtolength{\arraycolsep}{-3.5pt}
\begin{array}{lclclcl}
                                            &   &\DMatrix\rid{1}\VecNeq{W}\rid{1}& + &\EMatrix\rid{1}\VecNeq{W}\rid{2}  & = &\VecNeq{B}\rid{1}\\
\CMatrix\rid{k}\VecNeq{W}\rid{k-1}          & + &\DMatrix\rid{k}\VecNeq{W}\rid{k}& + &\EMatrix\rid{k}\VecNeq{W}\rid{k+1}& = &\VecNeq{B}\rid{k}\\
\CMatrix\rid{\Ngrid}\VecNeq{W}\rid{\Ngrid-1}& + &\DMatrix\rid{\Ngrid}\VecNeq{W}\rid{\Ngrid}&&                           & = &\VecNeq{B}\rid{\Ngrid}.
\end{array}
\addtolength{\arraycolsep}{3.5pt}
\ee
The algorithm operates in two passes. First, going from $k=1$ to
$\Ngrid$, one defines recursively a sequence of {\Neq}--vectors
$\VecNeq{V}\rid{k}$ and $({\Neq}\times {\Neq})$--matrices
$\MatNeq{M}\rid{k}$ through
\be
\begin{aligned}
\VecNeq{V}\rid{1} &= \left(\DMatrix\rid{1}\right)^{-1} \VecNeq{B}\rid{1}\\
\MatNeq{M}\rid{1} &= \left(\DMatrix\rid{1}\right)^{-1} \EMatrix\rid{1} \\
\VecNeq{V}\rid{k} &= \left(\DMatrix\rid{k}-\CMatrix\rid{k}\MatNeq{M}\rid{k-1}\right)^{-1} \left(\VecNeq{B}\rid{k}-\CMatrix\rid{k}\VecNeq{V}\rid{k-1}\right)\\
\MatNeq{M}\rid{k} &= \left(\DMatrix\rid{k}-\CMatrix\rid{k}\MatNeq{M}\rid{k-1}\right)^{-1} \EMatrix\rid{k}\qquad 2\le k \le \Ngrid.
\end{aligned}
\ee
$\MatNeq{M}\rid{\Ngrid}$ is not defined. In the second pass, the
values of the unknown $\VecNeq{V}\rid{k}$ are computed, climbing back
from $k=\Ngrid$ to $1$, with 
\be
\begin{aligned}
 \VecNeq{W}\rid{\Ngrid} &= \VecNeq{V}\rid{\Ngrid}\\
 \VecNeq{W}\rid{k} &= \VecNeq{V}\rid{k} - \MatNeq{M}\rid{k}\VecNeq{W}\rid{k+1}\qquad 1\le k \le \Ngrid-1.
\end{aligned}
\ee
Note that, with this algorithm, only $({\Neq}\times {\Neq})$ matrices
have to be inverted. We use Gauss elimination for this purpose because
this venerable technique proves to be robust enough to properly deal
with the kind of badly conditioned matrices that often appear in this
application.

The initial model for the Newton-Raphson algorithm is given by the
structure of the cluster at the previous time,
$\GrandVect{X}\itid{0}(t)=\GrandVect{X}(t-\Delta t)$ One iterates
until the following convergence criteria are met. Let us set $\delta
x\qid{i}\rid{k} \equiv
\left.x\qid{i}\rid{k}\right|\itid{m+1}-\left.x\qid{i}\rid{k}\right|\itid{m}$. Then,
the condition for logarithmic quantities is
\be
 \max_{i=1...\Neq} \frac{1}{\Ngrid} \sum_{k=1...\Ngrid} 
 \left(\delta x\qid{i}\rid{k}\right)^2  < \varepsilon_1,
\ee
with $\varepsilon_1=10^{-6}$. For velocities ($u$, $v_r-u$,
$v_t-u$), one checks
\be
\max_{i=1...\Neq} \frac{1}{\Ngrid} \sum_{k=1...\Ngrid} 
\left(\frac{ \delta x\qid{i}\rid{k} }{ x\qid{i}\rid{k}+\varepsilon_1 w\rid{k} }\right)^2  < \varepsilon_2,
\ee
with $\varepsilon_2=10^{-3}$ and $w\rid{k}=r\rid{k}(4\pi
G\rho\rid{k})^{1/2}$. Generally, two iterations are sufficient to
reach convergence.

\section{Velocity dispersion in the central regions}
\label{apdx:sigma}

\bfig 
\centerline{\resizebox{0.95\hsize}{!}{\includegraphics{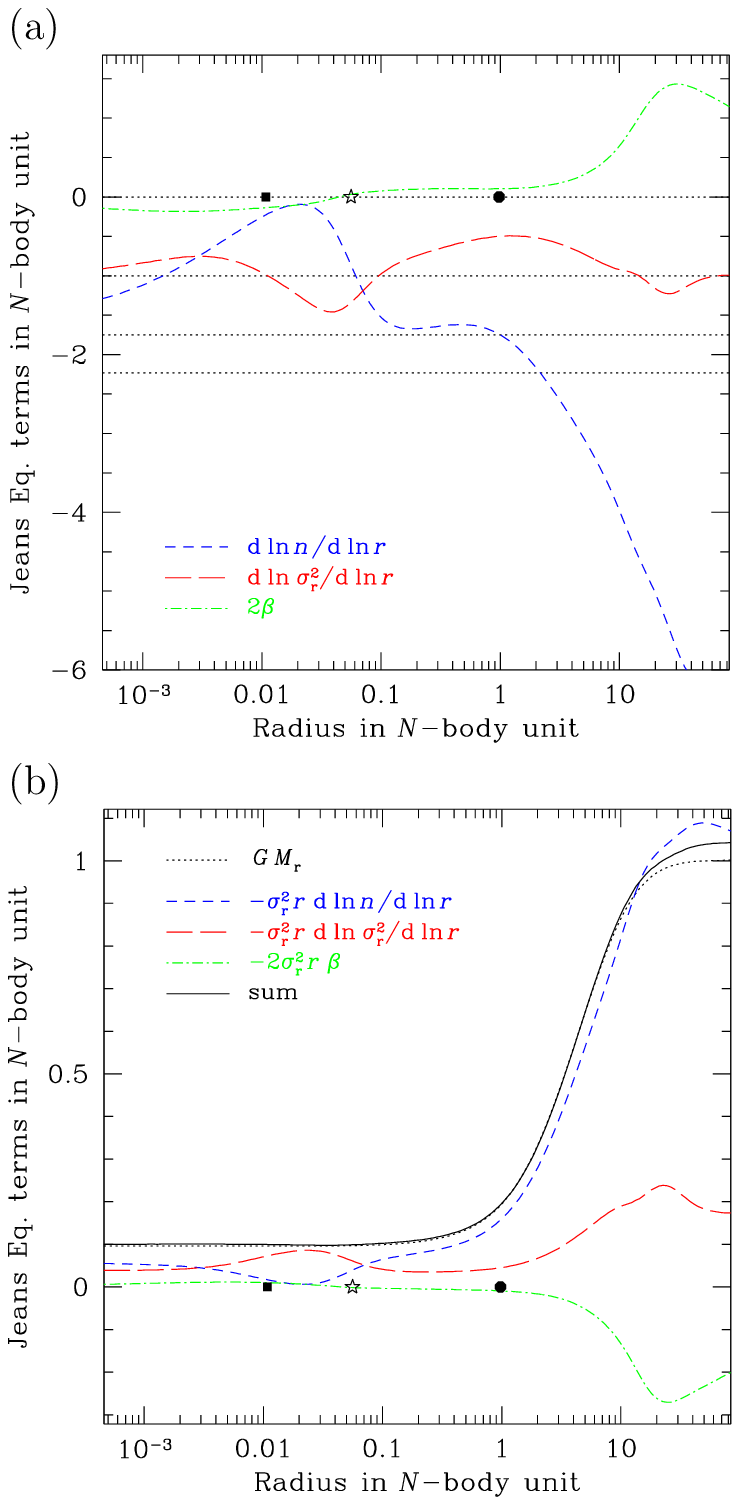}}}
\caption {
  Check of stationary Jeans equation for our ``standard'' model
  ($10^5$ stars, case a) at $T=10\,Gyrs$ (same model and time as
  Fig.\,\ref{fig.disp_a1e5_10Gyr}).  Panel (a) depicts the logarithmic
  derivatives of the stellar density $n$ and radial component of the
  velocity dispersion well as the anisotropy parameter,
  $2\beta=2-\sigma_t^2/\sigma_r^2$. Horizontal lines corresponding to
  values of $0$, $-1$, $-1.75$ and $-2.23$ are present to guide the
  eye. In particular, the ``Keplerian'' velocity profile is ${\rm
    d}\ln\,\sigma_r^2/{\rm d}\ln\,r\equiv -1$. The round dot indicates
  the influence radius, the star the critical radius and the square
  the ``1-star'' radius. On panel (b), we plot the three terms of the
  right side of the stationary Jeans equation and check that their sum
  is (nearly) equal to the left side term, i.e.\ $GM_{\rm r}$. 
  \label{fig.jeans}}
\efig 

\bfig 
\centerline{\resizebox{0.95\hsize}{!}{\includegraphics{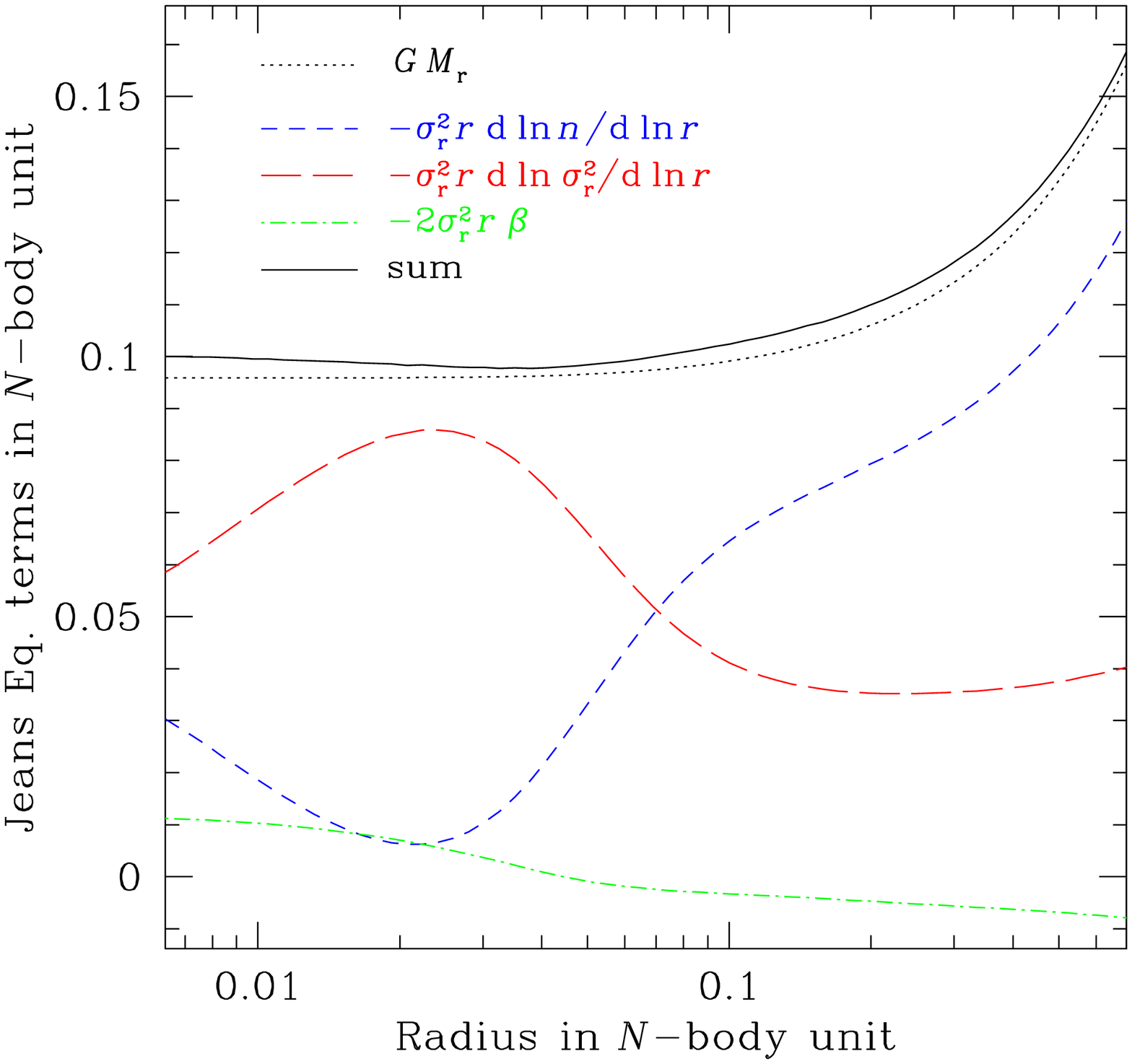}}}
\caption {
  Blow-up of panel (b) of Fig.\,\ref{fig.jeans} for the region between
  the 1-star and the influence radius. 
  \label{fig.jeansb}}
\efig 

In the region dominated by the central BH, one may expect a
``Keplerian'' profile for the velocity dispersion, $\sigma \propto
r^{-1/2}$. However, in Sec.\,\ref{sec.results}, we have seen that in our
standard model ($10^5$ stars, case a this relation does not really
apply where expected, i.e.\ between the ``1-star'' and the influence
radius. Here we show that the spherical Jeans equation for a system in
dynamical equilibrium is nevertheless obeyed.

In spherical symmetry the assumption of dynamical equilibrium (or
stationarity) amounts to $u = \left< v_{\rm r} \right> \equiv 0$. We
note that the gaseous model can cope with $u\neq 0$. On the other
hand, for a system whose evolution is driven by relaxation, one
expects $\sigma_{\rm r,t} \gg u$. The Jeans equation then reads
(\citealt{BT87}, Eq.\,4-55)

\be
GM_{\rm r} = -r\sigma_{\rm r}^2 \left(\frac{{\rm d}\ln n}{{\rm d}\ln r}
+\frac{{\rm d}\ln \sigma_{\rm r}^2}{{\rm d}\ln r}+2\beta\right).
\label{eq.jeans}
\ee $M_{\rm r}$ is the mass enclosed by the radius $r$, $n$ the number
density of stars and $2\beta=2-\sigma_t^2/\sigma_r^2$ is the
anisotropy parameter; other quantities have been defined previously.
One sees easily that if $M_{\rm r}\equiv \mbh$ and the first and third
term in the brackets are both constant, $\sigma \propto r^{-1/2}$ at
small $r$. But, as figures~\ref{fig.jeans} and \ref{fig.jeansb}
demonstrate, none of these assumptions exactly applies in the range of
radius under consideration. Consequently, we do not get a clean
Keplerian velocity profile although (or {\em because})
Eq.\,\ref{eq.jeans} is satisfied. Finally, we mention that our models
with $10^6$ and $10^7$ stars (and same initial size) exhibit a
Keplerian velocity cusp outside the 1-star radius, during their
post-collapse evolution. This is partly due to the relatively more
massive black hole (larger influence radius) and partly to the much
smaller 1-star radius.

\section{MBH wandering in a cuspy cluster.}
\label{apdx:wander}

\def\mcl{{\cal M}_{\rm cl}} 
\def\m{{\cal M}} 

Here we present a simple estimate of the wandering radius $R_{\rm
  wan}$ of a MBH embeded in a stellar cluster whose density posseses a
power-law cusp in the inner regions. We assume that, were it not for
the effect of the MBH itself, the stellar cluster would be described
by an eta-model \citep{Dehnen93,Tremaine94} with enclosed stellar mass
\be 
\m_\ast(r) = \mcl \left(\frac{r/a}{1+r/a}\right)^{\eta}, 
\ee
where $\mcl$ is the total mass in stars and $a$ the break radius. For
$r\ll a$, the density is $\rho \propto r^{-\alpha}$ with
$\alpha=3-\eta$. Inside $R_{\rm wan}$, the MBH strongly perturbates
stellar orbits and we suppose the density is rendered more or less
constant. Hence, the potential felt by the MBH is approximately
harmonic,
\be 
\Phi(r) = \Phi_0 + \frac{1}{2}\omega^2r^2\ \ \mbox{, with}\ 
\omega^2=\frac{G\m_\ast(R_{\rm wan})}{R_{\rm wan}^3}.  
\ee
For a harmonic oscillator, the RMS amplitude of the oscillations in
velocity and space are linked to each other,
\be
 V_{\rm RMS}^2 = \omega^2 R_{\rm RMS}^2 \approx \omega R_{\rm wan}^2 = 
\frac{G\m_\ast(R_{\rm wan})}{R_{\rm wan}}.
\label{eq:Vrms}
\ee
\citet{DHM03} have verified with $N$-body simulations that
equipartition of kinetic energy between the MBH and the stars is
established, at least in the case of $\eta=1.5$. Namely,
\be
\mbh V_{\rm RMS}^2 \simeq m_\ast \sigma^2,
\label{eq:Vrms2}
\ee
where $\sigma$ is the stellar velocity dispersion at $r=a$. For
$\eta=1.5$ \citep{Tremaine94},
\be 
\sigma^2 \simeq 0.1 \frac{G\mcl}{a}.
\label{eq:sigma}
\ee
Finally, assuming $R_{\rm wan} \ll a$ and, hence, $\m_\ast(R_{\rm
  wan}) \simeq \mcl \left(R_{\rm wan}/a\right)^\eta$ and combining
equations~\ref{eq:Vrms}, \ref{eq:Vrms2} and \ref{eq:sigma}, we obtain
\be
R_{\rm wan} \approx 0.01 a\left(\frac{m_\ast}{\mbh}\right)^2\ \ \mbox{for}\ \eta=1.5
\ee
and
\be
R_{\rm wan} \propto a\left(\frac{m_\ast}{\mbh}\right)^{1/(\eta-1)}
\ee
for general eta-models.

\section{Projected velocity dispersions}

For the Figures\,\ref{fig.projdisp_a1e5_10Gyr} and
\ref{fig.projdens_a1e5_10Gyr} we have integrated the density along the
z-axis for the projection,
\begin{eqnarray}
\lefteqn{
\Sigma (r)= \int_{z=0}^{z=R_{\rm max}} \rho(\sqrt{r^2+z^2})\,dz ={}} 
                                                 \nonumber \label{eq.sigma_proj} \\
& & {} 2 \int_{r}^{z={R_{\rm max}}} \rho(R)\frac{R}{\sqrt{R^2-r^2}}\,dR
\end{eqnarray} 

\noindent
If one observes the cluster along the z-axis, the contributions to the
projected velocity dispersions are as indicated in
Fig.\,\ref{fig.proj_vel_disp}. We have that $R=\sqrt{r^2+z^2}$ and
$\sigma_{\theta}= \sigma_{\phi}=\sigma_{\rm t}$, where the subscript t
stands for tangential. We can reckon that

\begin{eqnarray}
\sigma^2_{\rm z}=\sigma^2_{\rm R}\,\cos^2\theta + \myH{\sigma}^2_{\rm t}\sin^2 \theta \nonumber \\
\sigma^2_{\rm r}=\sigma^2_{\rm R}\,\sin^2\theta + \myH{\sigma}^2_{\rm t}\cos^2 \theta,
\end{eqnarray}

\noindent
where we have defined $\myH{\sigma}^2_{\rm t} \equiv \sigma^2_{\rm
  t}/2$, to be consistent with the notation used until now.

\bfig 
\resizebox{\hsize}{!}{\includegraphics[clip]{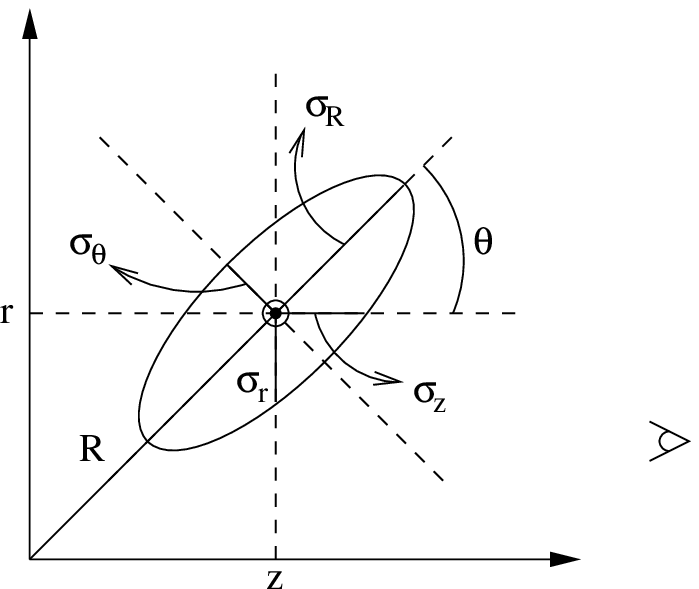}}
\caption{Different contributions to the projected velocity dispersions.
  The semi-major axis of the ellipsoid perpendicular to the page
  corresponds to $\sigma_{\rm t}=\sigma_{\phi}$.
\label{fig.proj_vel_disp} 
}
\efig

\noindent
In Fig.\,\ref{fig.proj_vel_disp} we can see that $\sigma_{\rm z}$
contributes to $\sigma_{\rm LOS}$, $\sigma_{\rm r}$ to $\sigma_{\rm
  pm,r}$ and $\myH{\sigma}_{\rm t}^2$ to $\sigma_{\rm pm,t}$.

Thus, we obtain the projected velocity dispersions,

\begin{eqnarray}
\lefteqn{
\sigma^2_{\rm LOS}(r)=\frac{2}{\Sigma(R)}\int^{R_{\rm max}}_{z=0}(\sigma^2_{\rm R}(R)\cos^2\theta+
\myH{\sigma}^2_{\rm t}(R)\sin^2\theta)\rho(R)\,dz={}}
\nonumber \\
&& {} \frac{2}{\Sigma(R)}\int^{R_{\rm max}}_{z=0}(\frac{z^2}{R^2}(\sigma^2_{\rm R}(R)-\myH{\sigma}^2_{\rm t}(R))+
\myH{\sigma}^2_{\rm t}(R))\rho(R)\,dz \nonumber
\end{eqnarray}
\begin{eqnarray}
\lefteqn{
\sigma^2_{\rm pm,r}(r)=\frac{2}{\Sigma(R)}\int^{R_{\rm max}}_{z=0}(\sigma^2_{\rm R}(R)\sin^2\theta+
\myH{\sigma}^2_{\rm t}(R)\cos^2\theta)\rho(R)\,dz={}} \nonumber \\ 
&& {} \frac{2}{\Sigma(R)}\int^{R_{\rm max}}_{z=0}(\frac{z^2}{R^2}(\myH{\sigma}^2_{\rm t}(R)-{\sigma}^2_{\rm R}(R))+
{\sigma}^2_{\rm R}(R))\rho(R)\,dz \nonumber
\end{eqnarray}
\begin{eqnarray}
\lefteqn{
\sigma^2_{\rm pm,r}(r)=\frac{2}{\Sigma(R)}\int^{R_{\rm max}}_{z=0}\myH{\sigma}^2_{\rm t}(R)\rho(R)\,dz,}
\label{eq.proj_vel_disp}
\end{eqnarray}

\noindent
since $\sin^2\theta=r^2/R^2=1-\cos^2\theta$ and $\cos^2\theta=z^2/R^2$.

\section*{Acknowledgements}
 
This work has been supported by Sonderforschungsbereich (SFB) 439
``Galaxies in the Young Universe'' of German Science Foundation (DFG)
at the University of Heidelberg, performed under the frame of the
subproject A5. PAS would like to thank Els Etsus for the
fruitful e-mailing.

\bibliographystyle{mn}
\bibliography{aamnem99,biblio,bib_marc,bib_pau}
\label{lastpage}
\end{document}